\documentclass{article}

\usepackage[utf8]{inputenc}
\usepackage{amsmath,amssymb,hyperref}
\usepackage{cite,braket}
\usepackage{titlesec}
\usepackage{simpler-wick}
\usepackage{graphicx}
\usepackage{tikz-feynman}\tikzfeynmanset{compat=1.1.0}

\allowdisplaybreaks

\def \ra {\rangle}
\def \la {\langle}

\def \be {\begin{equation}}
\def \ee {\end{equation}}
\def \ba {\begin{eqnarray}}
\def \ea {\end{eqnarray}}

\def \Dfree {D_{\rm free}}
\def \Dloop {D_{\rm one\!-\! loop}}
\def \ph {\phantom{\frac11}}

\usepackage{color}

\title{\bf\boldmath Energy-momentum tensor in $\Phi^4$ theory at one loop}

\author{Brean Maynard\\
  \footnotesize Department of Physics, University of Connecticut, Storrs, CT 06269, U.S.A.\\
  \footnotesize May 2024}

\date{ } 

\begin{document}

\maketitle

\begin{abstract}
The energy-momentum tensor form factors are studied in $\Phi^4$ theory to one-loop 
order with particular focus on the $D$-term, a particle property which has attracted a lot of 
attention in the recent literature. It is shown that the free Klein-Gordon theory value of the 
$D$-term $\Dfree=-1$ is reduced to $\Dloop =-\frac13$ even if the $\Phi^4$ interaction
is infinitesimally small. 
A companion work in $\Phi^3$ theory confirms this result which may indicate that it is
independent of the type of interaction as long as the scalar theory is renormalizable. 
Dispersion relations are studied. Various definitions of mean square radii including the
mechanical radius are investigated. The findings contribute to a better understanding 
of the EMT properties of particles and their interpretation.
\end{abstract}

\section{Introduction}
\label{Sec-1:intro}

The energy momentum tensor (EMT) is an important operator in field theory through
which particles couple to gravity, and whose matrix elements define fundamental 
particle properties like mass, spin and the less well-known $D$-term. Introduced 
in the 1960s \cite{Kobzarev:1962wt,Pagels:1966zza}, EMT form factors received little 
attention until the advent of generalized parton distribution functions accessible 
in hard-exclusive reactions 
\cite{Muller:1998fv,Ji:1996ek,Radyushkin:1996nd,Radyushkin:1996ru,Ji:1996nm,Collins:1996fb}
which allow one to infer the spin \cite{Ji:1996ek} and mass\cite{Ji:1994av} decompositions 
of the nucleon, the property known as the $D$-term \cite{Polyakov:1999gs},
and mechanical properties \cite{Polyakov:2002yz}. For reviews on 
generalized parton distribution functions and EMT form factors, see 
\cite{Ji:1998pc,Goeke:2001tz,Diehl:2003ny,Belitsky:2005qn,Boffi:2007yc,Guidal:2013rya,
dHose:2016mda,Kumericki:2016ehc,Polyakov:2018zvc,Burkert:2023wzr}. 

In view of the complexity of hadron structure and QCD, it is interesting to study the
EMT properties of particles in simpler theories. The simplest case constitute free theories 
where EMT matrix elements and other properties can be exactly evaluated. For instance, 
the $D$-term of a free, non-interacting spin-zero particle is $D_{\rm free}=-1$ \cite{Hudson:2017xug}, 
while that of a free fermion vanishes \cite{Hudson:2017oul}. 
The arguably ``next-to-simplest'' case are theories with an infinitesimally small interaction
which can be solved perturbatively to lowest order. 

The purpose of this work is to investigate how the EMT form factors are affected if 
the theory of a scalar field $\Phi$ is introduced to a $\Phi^4$ type interaction. 
To be more precise, the investigation will be carried out in the perturbative regime 
under the assumption that the coupling constant $\lambda \ll 1$ is ``infinitesimally small''
such that a one-loop calculation can be assumed to be justified and higher orders 
${\cal O}(\lambda^2)$ can be safely neglected. The goal of this work is to determine the
EMT form factors to one-loop order in the $\Phi^4$ theory.

At the first glance, in the situation of an infinitesimally small coupling constant in a 
scalar theory, one could naively expect the one-loop result $\Dloop$ to be not much different 
from the free-scalar theory result $\Dfree=-1$. However, this is not the case. Rather, the 
$D$-term is strongly affected when interactions are introduced in the theory of a scalar
field, and its value is reduced at one-loop order to $\Dloop = -\frac13$. This was suspected
(but not proven) in Ref.~\cite{Hudson:2017xug}. The technical reason for this strong shift of 
the value of the $D$-term is due to the necessity to introduce an improvement
term to render Greens functions with EMT insertions finite \cite{Callan:1970ze}, see also 
\cite{Coleman:1970je,Freedman:1974gs,Freedman:1974ze,Lowenstein:1971vf,Schroer:1971ud,Collins:1976vm}.

Interestingly, the same finding was obtained in another renormalizable scalar theory, 
namely in $\Phi^3$ theory in \cite{Dotson:new}. This may indicate that the $D$-term is 
affected in the same way by infinitesimal interactions, independently of the type of
interaction as long as it is renormalizable. 
These findings are in line with observations in literature that the $D$-term is the particle
property most sensitive to the variations of the dynamics in a system  
\cite{Polyakov:2018zvc,Burkert:2023wzr,Hudson:2017xug,Hudson:2017oul}.

The outline is as follows. 
In Sec.~\ref{Sec-2:EMT-spin-zero} the EMT form factors for spin-zero particles are presented,  
free-field theory results reviewed, and $\Phi^4$ theory is introduced.
In Secs.~\ref{Sec-3:EMT-dim-reg} and \ref{Sec-4:EMT-PV} the one-loop calculation is
carried out in respectively dimensional and Pauli-Villars regularization.
In Sec.~\ref{Sec-5:properties-of-D(t)} the properties of the $D(t)$ form factor
are discussed, and Sec.~\ref{Sec-6:dispersion-relation} presents a non-trivial
cross check of the calculation by means of dispersion relations.
The Sec.~\ref{Sec-7:radii} discusses several mean square radii.
In Sec.~\ref{Sec-8:conclusions} the conclusions are presented. 
The Appendices contain technical details.

\newpage
\section{EMT form factors of a spin-zero particle}
\label{Sec-2:EMT-spin-zero}

In this section, we introduce the EMT form factors for a spin-zero particle,
review what is known from the free theory case and discuss expectations when 
interactions are included.

\subsection{EMT form factors}

The matrix elements of the EMT operator of a spin-zero particle are 
described in terms of 2 form factors as \cite{Kobzarev:1962wt,Pagels:1966zza}
\be\label{Eq:def-EMT-FF}
	\la\vec{p}^{\,\prime\,}|T^{\mu\nu}(x)|\vec{p}\ra = 
	e^{i\Delta\cdot x}\biggl(\frac{P^\mu P^\nu\!}{2}\, A(t) + 
	\frac{\Delta^\mu\Delta^\nu - g^{\mu\nu}\Delta^2}{2}\,D(t)\biggr)
\ee
where $T^{\mu\nu}(x)$ denotes the EMT operator at the space-time 
position $x^\mu$, and the variables are defined as
\be\label{Eq:define-variables}
	P^\mu=p^{\,\prime\,\mu}+p^\mu\, , \;\;\;
	\Delta^\mu=p^{\,\prime\,\mu}-p^\mu\, , \;\;\; t = \Delta^2.
\ee
The covariant normalization of the one-particle states is provided by
\be\label{Eq:norm-states}
	\la\vec{p}^{\,\prime}|\vec{p}\,\ra = 2\,\omega_p\,(2\pi)^3\,
	\delta^{(3)}(\vec{p}-\vec{p}^{\,\prime})\,, \quad
	\omega_p=\sqrt{\vec{p}^{\,\,2}+m^2}\,.
\ee
At zero-momentum transfer, the form factors satisfy
\cite{Kobzarev:1962wt,Pagels:1966zza}, see for instance also 
Ref.~\cite{Cotogno:2019xcl}, 
\be\label{Eq:constraint-EMT-FF}
	\lim\limits_{t\to 0} A(t) = A(0) = 1\,, \quad
	\lim\limits_{t\to 0} D(t) = D(0) \equiv D \,.
\ee
The constraint for $A(0)=1$ arises because in the limit $\vec{p}\to 0$ and
$\vec{p}^{\,\prime}\to 0$, only the $00$-component contributes in 
Eq.~(\ref{Eq:def-EMT-FF}), and $H=\int d^3x\;T^{00}(x)$ is 
the Hamiltonian of the system with $H\,|\vec{p}\ra = m\,|\vec{p}\ra$ 
when $\vec{p}\to 0$.
It is important to notice that no analog constraint exists for $D(0)$
which therefore implies a novel global property of a particle
which is on the same footing as other fundamental particle properties 
related to matrix elements of operators corresponding to conserved 
currents like mass, spin, electric charge or magnetic moment 
\cite{Polyakov:1999gs}.

Notice that many authors define the vector $P^\mu$ not as a sum 
of $p^\mu$ and $p^{\prime\mu}$ but their average. Due to invariance under space-time
translations, the position $x^\mu$ of the EMT operator can be arbitrarily changed 
which affects only the overall phase $e^{i\Delta\cdot x}$ in Eq.~(\ref{Eq:def-EMT-FF}). 
The position $x^\mu$ can therefore be set to the origin without loss of generality.

\subsection{Free Klein-Gordon theory}

The classical Lagrangian of the theory of a real spin-zero particle of 
mass $m$ is given by
\be\label{Eq:L-free}
    \mathcal{L}_{\rm free}=\frac{1}{2}\,(\partial^\rho\Phi)\,(\partial_\rho\Phi)-\frac{1}{2}m^2\Phi^2 \,.
\ee
By exploring translational invariance of the theory, the Noether theorem yields the canonical EMT 
which is symmetric and given by the expression
\be\label{Eq:EMT-free}
    T^{\mu\nu}_{\rm free}(x) = (\partial^{\mu}\Phi)\,(\partial^\nu \Phi)-g^{\mu\nu}\mathcal{L}_{\rm free} \,.
\ee
The Euler-Lagrange equation of the free theory in Eq.~(\ref{Eq:L-free}) is the Klein-Gorden equation
$(\square+m^2)\Phi(x)=0$ with the free field solution given by
\begin{equation}
    \Phi(x) =\int\frac{d^3k}{2\omega_{k} (2\pi)^3}
    \biggl(\hat{a}_{k}e^{-ik\cdot x}+\hat{a}^{\dagger}_{k}e^{ik \cdot x}\biggr) \,
\end{equation}
where $[\hat{a}_k,\hat{a}_{k'}]=2\omega_k(2\pi)^3\delta^3({\vec{k}-\vec{k'}})$ and 
$\omega_k$ is defined in Eq.~(\ref{Eq:norm-states}). The one-particle states are given by 
$\ket{\vec{k}}=\hat{a}^{\dagger}_{k}\ket{0}$, where $\ket{0}$ is the free theory vacuum state.
Evaluating the matrix elements of $T^{\mu\nu}$ for $\lambda=0$ yields for the form factors
\begin{equation} \label{Eq:EMT-FFs-free}
    A(t)_{\rm free} =  1\,, \quad
    D(t)_{\rm free} = -1\,.
\end{equation}
These results were first obtained in Ref.~\cite{Pagels:1966zza}, see also Ref.~\cite{Hudson:2017xug}.

\subsection{\boldmath The $\Phi^4$ theory}

We consider the theory of a real scalar particle of mass $m$ with a self-interaction $\Phi^4$ 
and the coupling constant $\lambda$. The classical Lagrangian of this theory is given by
\be\label{Eq:L}
    \mathcal{L}=\frac{1}{2}\,(\partial^\rho\Phi)\,(\partial_\rho\Phi)-\frac{1}{2}m^2\Phi^2 
    - \frac{\lambda}{4!}\,\Phi^4 \,.
\ee
The EMT of this theory is given by the expression
\be\label{Eq:EMT}
    T^{\mu\nu}(x)
    =T^{\mu\nu}_{\rm can}(x) + \Theta^{\mu\nu}_{\rm imp}(x)\,, \quad
    T^{\mu\nu}_{\rm can}(x) = (\partial^{\mu}\Phi)\,(\partial^\nu \Phi)-g^{\mu\nu}\mathcal{L}
\ee
where the first term $T^{\mu\nu}_{\rm can}(x)$ is the symmetric canonical EMT which follows 
from the Noether theorem, while $\Theta^{\mu\nu}_{\rm imp}(x)$ constitutes the improvement 
term which is given in $N$ space-time dimensions by \cite{Callan:1970ze}
\be\label{Eq:EMT-improve}
	\Theta^{\mu\nu}_{\rm imp}(x) = -h
	(\partial^\mu\partial^\nu-g^{\mu\nu}\square)\,\Phi^2(x)  \,,
	\;\;\;\;\;
	h = \frac14\biggl(\frac{N-2}{N-1}\biggr) ,
\ee
with $h=\frac16$ in $N=3+1$ space-time dimensions.
The improvement term (\ref{Eq:EMT-improve}) can be derived from coupling the scalar
theory (\ref{Eq:L}) to a classical gravitational background field in a non-minimal 
way in terms of the effective action
\be\label{Eq:S-grav-II}
	S_{\rm grav} = \int d^4x\;\sqrt{-g}\biggl(
	\frac12\,g^{\mu\nu}(x)(\partial_\mu\Phi)(\partial_\nu\Phi)- V(\Phi)
	-\frac12\,h\,R\,\Phi^2\biggr)
\ee
where $-\frac12\,h\,R\,\Phi^2$ is a non-minimal coupling term, $R$ is the Riemann scalar, 
and $g$ denotes the determinant of the space-time dependent metric $g^{\mu\nu}(x)$.
In this work, $V(\Phi)=\frac12m^2\Phi^2+\frac{1}{4!}\lambda\Phi^4$,
but the improvement term is valid also for other scalar theories.
The EMT operator in Eqs.~(\ref{Eq:EMT},~\ref{Eq:EMT-improve}) follows from varying
the action (\ref{Eq:S-grav-II}) according to
\be\label{Eq:EMT-from-gravity-II}
	T^{\mu\nu}(x) = \frac{2}{\sqrt{-g}}\,
	\frac{\delta S_{\rm grav}}{\delta g_{\mu\nu}(x)}\,,
\ee
where it is understood that after the variation, the metric is set to the metric of the flat space.
Notice that the Riemann scalar $R$ vanishes in flat space, but its variation 
with respect to the metric is nevertheless non-zero.

On the quantum level, the addition of the improvement term makes Greens functions of 
renormalized fields with an EMT insertion finite \cite{Callan:1970ze}. To be more precise, 
the $h$ quoted in Eq.~(\ref{Eq:EMT-improve}) removes UV divergences up to three-loops in dimensional 
regularization in $\Phi^4$ theory. The operator expression for the improvement term remains the same 
to all orders, albeit the value of $h$ needs to be changed beyond three loops \cite{Collins:1976vm}. 

The coupling constant $\lambda\ll 1$ can be thought to be infinitesimally small. In the $\Phi^4$ theory,
the cross section for the scattering of the $\Phi$-particles is proportional to $\lambda^2$. 
Clearly, for an infinitesimally small $\lambda\ll 1$ the cross section is negligibly small and 
the $\Phi$-particles are practically free. In that case, one could naively assume that the
EMT form factors hardly change compared to the free theory result in Eq.~(\ref{Eq:EMT-FFs-free})
\cite{Hudson:2017xug,Ji:2021mfb}.
But this is not the case, as it was suspected (without proof) in Ref.~\cite{Hudson:2017xug} 
and will be explicitly demonstrated in the subsequent sections.

\section{\boldmath EMT in $\Phi^4$ theory at one-loop in dimensional regularization}
\label{Sec-3:EMT-dim-reg}

To one-loop order under renormalized perturbation theory, the Lagrangian in Eq.~(\ref{Eq:L})
becomes
\begin{equation}\label{Eq:L-dim-reg}
    \mathcal{L}=\frac{1}{2}(\partial_\rho\Phi)(\partial^\rho\Phi)-\frac{1}{2}(m^2+\delta_m)\Phi^2-\frac{\lambda \Phi^4}{4!}.
\end{equation}
The mass counter term is set by the on-shell renormalization conditions,
\begin{equation}
   \left.M^2\left(p^2\right)\right|_{p^2=m^2}=0 \quad \text { and }\left.\quad \frac{d}{d p^2} M^2\left(p^2\right)\right|_{p^2=m^2}=0\label{onshell1}
\end{equation}
where $-iM^2(p^2)$ is the sum of one-particle-irreducible (1PI) insertions. With these conditions, $ \delta_{m}$ is set to one-loop order by 
\begin{equation}
    -iM^2(p^2)=-i\mu^{4-d} \lambda \;\frac{1}{2} \; \int \frac{d^d k}{(2 \pi)^d} \frac{i}{k^2-m^2}+i\left(p^2 \delta_Z-\delta_m\right)\label{onshell2},
\end{equation}
where the wave function renormalization constant $\delta_Z=0$ at one loop. 
In dimensional regularization, Eq.~\eqref{onshell2} implies 
 \be\label{Eq:counter-m-dim}
    \delta_{m}= -\lambda\frac{\mu^{4-d}}{2(4\pi)^\frac{d}{2}}\;\frac{\Gamma(1-\frac{d}{2})}{m^{2-d}}.
\ee
The coupling constant counterterm, $\delta_{\lambda}=0$ 
at the order $\lambda$ and becomes non-zero only at 
${\cal O}(\lambda^2)$, which is not considered in this work.

\subsection{\boldmath Canonical EMT in dimensional regularization}
\label{Sec-3.1:EMT-dim-reg-canonical}

Let us begin with the matrix elements of the canonical EMT which are given by
\ba\label{Eq:EMT-can-dim}
     \bra{ \vec{p}^{\:\prime}_{\rm out}} T^{\mu\nu}_{\rm can}(0) \ket{ \vec {p}_{\rm in}} 
     &=& 
     \bra{\Omega_{\rm out}}\hat{a}_{p'}(\partial^{\mu}\Phi\partial^{\nu} \Phi
     -g^{\mu \nu}\mathcal{L})\hat{a}^{\dagger}_p)\ket{\Omega_{\rm in}} \nonumber\\
     &=&
     \bra{\Omega}T\biggl(\hat{a}_{p'}(t')(\partial^{\mu}\Phi\partial^{\nu} \Phi
     -g^{\mu \nu}\mathcal{L})\hat{a}^{\dagger}_p(t) \,e^{-i\int \frac{\lambda}{4!}\Phi^4(y)d^4y}\biggr)\ket{\Omega}
     \;,
\ea
where $\ket{\Omega}$ denotes the vacuum of the interacting theory. In the last step
the distinction between in- and out-vacuum states is irrelevant and the limits $t\to-\infty$
and $t'\to\infty$ are understood.
Here $\Phi$ is redefined to be the renormalized wave function, and we will only calculate 
the connected, amputated contributions. Expanding to ${\cal O}(\lambda)$ yields
\ba\label{Eq:EMT-can-1}
    \bra{ \vec{p}^{\:\prime}_{\rm out}} T^{\mu\nu}_{\rm can}(0) \ket{ \vec {p}_{\rm in}}
    &=&\frac{P^{\mu}P^\nu}{2}-\frac{\Delta^{\mu}\Delta^\nu-g^{\mu \nu}\Delta^2}{2}
    +\lambda\biggl[f^{\mu\nu}(\Delta)
    +g^{\mu\nu}\biggl( g(\Delta)+
    m^2\,h(\Delta)+j(\Delta) 
    )\biggr)\biggr] 
    + g^{\mu\nu}\,\delta_m\ 
 \ea

\begin{figure}[b!]
\begin{centering}
\includegraphics[width=4.2cm]{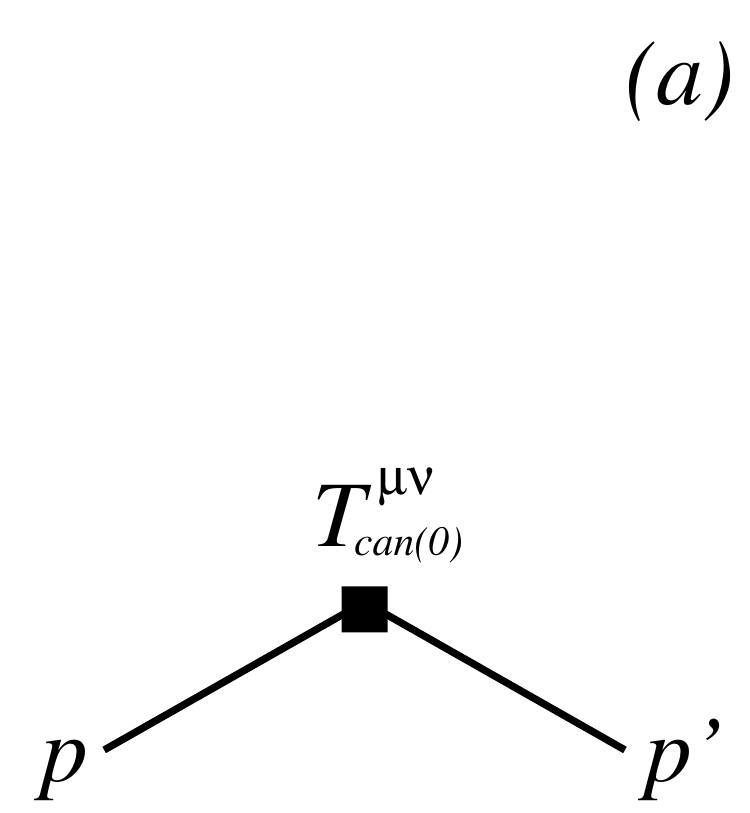} \hspace{2cm}
\includegraphics[width=4.2cm]{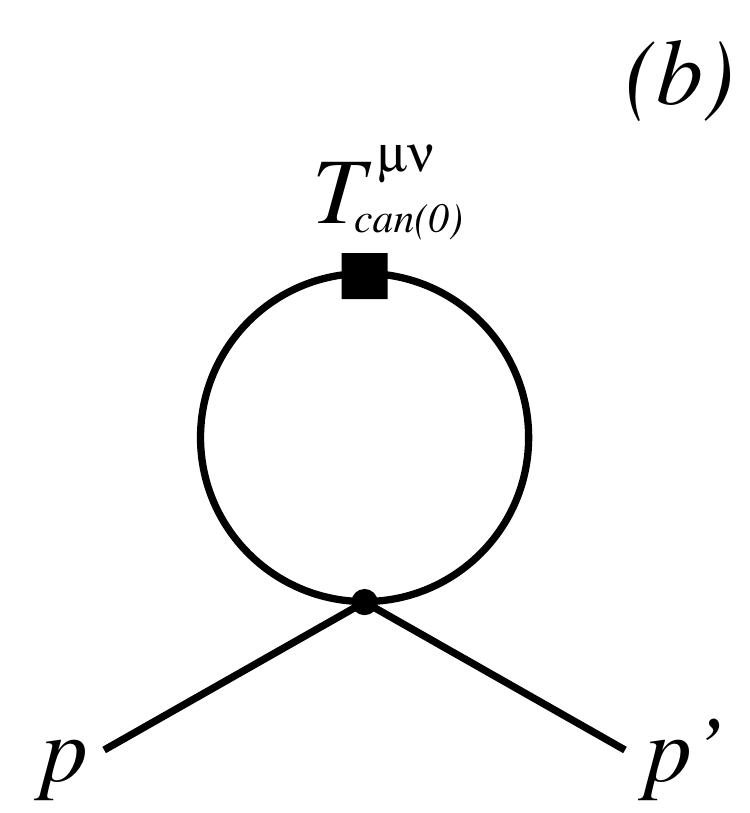} \hspace{2cm}
\includegraphics[width=4.2cm]{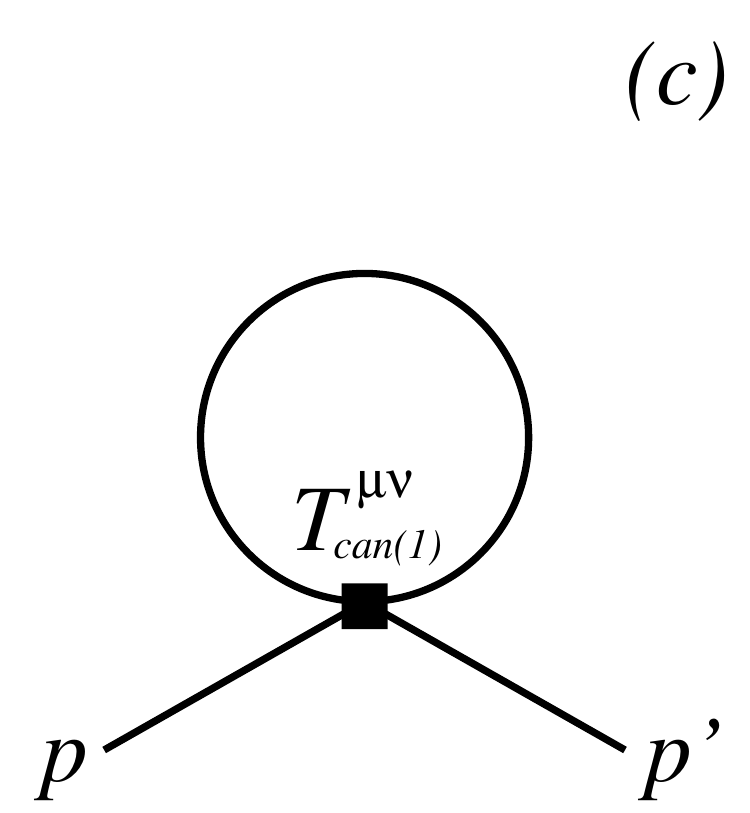} 
\par\end{centering}
\caption{\label{Fig:Diagrams} 
Feynman diagrams for the evaluation of matrix elements of $T^{\mu\nu}_{{\rm can}(n)}$ in $\Phi^4$ theory. 
The squares represent insertions of $T^{\mu\nu}_{{\rm can}(n)}$ at respectively $n=0$ order in $\lambda$ 
(with $T^{\mu\nu}_{{\rm can}(0)}=T^{\mu\nu}_{\rm free}(x)$ as defined in Eq.~(\ref{Eq:EMT-free}))
and $n=1$ order (where $T^{\mu\nu}_{{\rm can}(1)}=g^{\mu\nu}\frac{\lambda}{4!}\Phi^4$ 
which can be traced back to the operator $-g^{\mu\nu}{\cal L}$ in $T^{\mu\nu}_{\rm can}$ in Eq.~(\ref{Eq:EMT})).}
\end{figure}

The first two terms of ${\cal O}(\lambda^0)$ correspond to the tree-level diagram 
in Fig.~\ref{Fig:Diagrams}a. The terms $f^{\mu\nu}(\Delta)$, $g(\Delta)$, $h(\Delta)$ 
are due to insertions of respectively the operators $\partial^\mu\Phi\partial^\nu\Phi$,
$\partial_\rho\Phi\partial^\rho\Phi$ and $\Phi^2$ from the zeroth-order canonical EMT
and correspond to the loop diagram in Fig.~\ref{Fig:Diagrams}b where the vertex originates 
from the expansion of $e^{-i\int \frac{\lambda}{4!}\Phi^4(y)d^4y}$ to 
first order in $\lambda$. 
The term $j(\Delta)$ originates from $\frac{\lambda}{4!}\Phi^4$ in the canonical EMT 
in conjunction with the zeroth order of $e^{-i\int \frac{\lambda}{4!}\Phi^4(y)d^4y}$
and corresponds to the diagram in  Fig.~\ref{Fig:Diagrams}c.
The expressions $f^{\mu\nu}(\Delta)$, $g(\Delta)$, $h(\Delta)$, $j(\Delta)$ are logarithmically or quadratically 
divergent in $d=4$ space-time dimensions, and need to be regularized (indicated by the subscript ``reg'' below)
and are given by, cf.\ App.~\ref{App:A}, 
 \begin{alignat}{3}
    f^{\mu\nu}(\Delta) \ &=& i &\int_{\rm reg} \frac{d^4 k}{(2\pi)^4} \frac{k^{\mu}}{k^2-m^2+i\epsilon}\;
    \frac{(k-\Delta)^{\nu}}{(k-\Delta)^2-m^2+i\epsilon} \,, \nonumber\\
    g(\Delta) \ &=& -\frac{i}{2} &\int_{\rm reg}  \frac{d^4 k}{(2\pi)^4} \frac{k_{\rho}}{k^2-m^2+i\epsilon}\;
    \frac{(k-\Delta)^{\rho}}{(k-\Delta)^2-m^2+i\epsilon} \,, \nonumber\\
    h(\Delta) \ &=& \ \frac{i}{2} &\int _{\rm reg} \frac{d^4 k}{(2\pi)^4} \frac{1}{k^2-m^2+i\epsilon}\;
    \frac{1}{(k-\Delta)^2-m^2+i\epsilon} \, , \nonumber\\
    j(\Delta) \ &=&\frac{i}{2}&\int_{\rm reg} \frac{d^4 k}{(2\pi)^4}\frac{1}{k^2-m^2+i\epsilon} \,.
    \label{Eq:EMT-can-2}
 \end{alignat}

In dimensional regularization in $\Phi^4$ Theory, one replaces $d^4k/(2\pi)^4\to \mu^{d-4} d^dk/(2\pi)^d$ and 
introduces the renormalization scale $\mu$. After exploring the Feynman parameterization
$$
    \frac{1}{U\,V} = \int_0^1dx\,\frac{1}{(x\,U+(1-x)\,V)^2} \,,
$$
and carrying out a Wick rotation which results in $ik^0 \to k_E^0$, integrating over Euclidean
momenta yields, see App.~\ref{App:B},
 \ba
    f^{\mu\nu}(\Delta) 
    &=& 
    \frac{\mu^{4-d}}{2(4\pi)^{\frac{d}{2}}} \int_0^1 dx\biggl(g^{\mu\nu}\frac{\Gamma(1-\frac{d}{2})}{\overline{m}(x,t)^{2-d}}+
    \Delta^{\mu}\Delta^{\nu}(1-x)x\frac{2\,\Gamma(2-\frac{d}{2})}{\overline{m}(x,t)^{4-d}}\biggr)
    \nonumber\\
    g(\Delta)
    &=& 
    \frac{\mu^{4-d}}{2(4\pi)^{\frac{d}{2}}}  \int_0^1 dx\biggl(-\frac{\Gamma(1-\frac{d}{2})\,d}{\overline{m}(x,t)^{2-d}}
    -\Delta^2(1-x)x\frac{2\,\Gamma(2-\frac{d}{2})}{\overline{m}(x,t)^{4-d}}\biggr)
    \nonumber\\
    h(\Delta)
    &=& \frac{\mu^{4-d}}{2(4\pi)^{\frac{d}{2}}} \int_0^1 dx\biggl(-\,\frac{\Gamma(2-\tfrac{d}{2})}{ \overline{m}(x,t)^{4-d}}\biggr)
    \nonumber\\
    j(\Delta)
    &=& 
    \frac{\mu^{4-d}}{2(4\pi)^\frac{d}{2}}\;\frac{\Gamma(1-\frac{d}{2})}{m^{2-d}}
 \ea
where 
\be\label{Eq:OVERLINEm}
    \overline{m}(x,t)^2=m^2-x(1-x)\,t\,.
\ee 
Inserting these results in Eq.~(\ref{Eq:EMT-can-dim}), 
setting the number of space-time dimensions to $d=4-\epsilon$ and expanding in $\epsilon$
yields the following contributions to the EMT form factors from $T^{\mu\nu}_{\rm can}(0)$
\be\label{Eq:EMT-can-dim-result}
    A(t)_{\rm one\!-\!loop,\,can} =  1 \, , \quad
    D(t)_{\rm one\!-\!loop, \,can} = -1+\frac{2\lambda}{(4\pi)^2}\,
    \int_0^1 dx(1-x)x\biggl(\frac{2}{\epsilon}+\ln{\frac{\mu^2}{\overline{m}(x,t)^2}}\biggr) \,,
\ee
where it is understood that $\epsilon>0$ is small but non-zero and ${\cal O}(\epsilon)$
neglected. Several comments are in order. 

First, the canonical EMT has the Lorentz decomposition in Eq.~(\ref{Eq:def-EMT-FF}) and is therefore
evidently conserved. Second, the form factor $A(t)$ is not altered at one-loop as compared to the 
free theory case. Third, it is important to stress that the one-loop result for $A(t)$ complies 
with the general requirement $A(0)=1$ in Eq.~(\ref{Eq:constraint-EMT-FF}) which is a cross check 
for the consistency of the calculation. 
Fourth, the form factor $D(t)$ obtained in the one-loop calculation is divergent and depends
on the renormalization scale $\mu$. 

The latter observation reflects the known fact that in order to obtain a finite result for the
EMT matrix elements, it is necessary to include the improvement term \cite{Callan:1970ze}.
We expect that this will remedy both problems, namely make $D(t)$ (i) finite 
and (ii) renormalization scale independent without spoiling EMT conservation and the 
constraint $A(0)=1$. We will show this explicitly in Sec.~\ref{Sec-3.2:EMT-dim-reg-improve}.

\subsection{Including the improvement term in dimensional regularization}
\label{Sec-3.2:EMT-dim-reg-improve}

Evaluating the matrix elements of the improvement term introduced in Eqs.~(\ref{Eq:EMT},~\ref{Eq:EMT-improve})
in $\Phi^4$ theory yields a tree-level contribution independent of $\lambda$ and a loop contribution 
proportional to $\lambda$. The results are given as 
\be
    \bra{ \vec{p}^{\:\prime}_{\rm out}} \Theta^{\mu\nu}_{\rm imp}(0) \ket{ \vec {p}_{\rm in}} 
    = \frac{\Delta^{\mu}\Delta^\nu-g^{\mu\nu}\Delta^2}{2}\,4h\,\biggl(1
    +\lambda\,h(\Delta)\biggr) 
\ee
with the integral $h(\Delta)$ as defined in Eq.~(\ref{Eq:EMT-can-2}).
The improvement term generates the structure $(\Delta^\mu\Delta^\nu-g^{\mu\nu}\Delta^2)$ 
which preserves EMT conservation and contributes only to $D(t)$. Setting $d=4-\epsilon$, expanding 
in $\epsilon$ and neglecting ${\cal O}(\epsilon)$-terms, the contribution to $D(t)$ is 
\be\label{Eq:EMT-imp-dim-result}
    D(t)_{\rm one\!-\!loop,\,imp} 
    = 4h \biggl[1 - \frac{\lambda}{2(4\pi)^2}\biggl(\frac{2}{\epsilon}+\ln{\frac{\mu^2}{\overline{m}(x,t)^2}}\biggr) 
    \biggr]\,,
\ee
Setting $N=4$ space-time directions in Eq.~(\ref{Eq:EMT-improve}), which implies $h=\frac16$, and combining
the results (\ref{Eq:EMT-can-dim-result},~\ref{Eq:EMT-imp-dim-result}) to determine the EMT form 
factors of the total EMT operator $T^{\mu\nu}(x)=T^{\mu\nu}_{\rm can}(x) + \Theta^{\mu\nu}_{\rm imp}(x)$
at one-loop order in $\Phi^4$ theory yields
\be\label{Eq:EMT-tot-dim-result}
    A(t)_{\rm one\!-\!loop} =  1 \, , \quad
    D(t)_{\rm one\!-\!loop} = -\frac{1}{3}
    +\frac{2\lambda}{(4\pi)^2}\int_0^1 dx\biggl(\frac{1}{6}-(1-x)x\biggr)
    \log\biggr(1-\frac{t}{m^2}x(1-x)\biggl) \,.
 \ee
As expected, the addition of the improvement term has preserved the EMT conservation
and the constrained $A(0)=1$ (in fact, it left the free theory result 
$A_{\rm free}(t)=1$ unaffected). But it has affected $D(t)$ in the way it was expected,
namely making it finite and renormalization scale independent (we recall that the running
of the coupling constant begins at two loops, and $\lambda$ is renormalization scale independent
at one-loop order to which we work here).

Notice that throughout Sec.~\ref{Sec-3:EMT-dim-reg}, we have notationally distinguished 
the number of space-time dimensions of the physical space $N=4$ in Eq.~(\ref{Eq:EMT-improve}) 
and $d=4-\epsilon$ in the integrals requiring regularization. This careful distinction was deliberate
and necessary.
In Eq.~(\ref{Eq:EMT-improve}) the number of space dimensions $N=4$ must be kept fixed and must 
not be continued to $4-\epsilon$ in dimensional regularization. This can be understood in two ways.

First, had we continued $N=4$ to $4-\epsilon$ also in $h$ in Eq.~(\ref{Eq:EMT-improve}), 
then we would still obtain a finite one-loop result for $D(t)_{\rm one\!-\!loop}$ but the
$\mu$-dependence would not cancel out. 
However, the form factors of a conserved external current must be renormalization scale independent. 
Second, the improvement term in Eq.~(\ref{Eq:EMT-improve}) was not constructed with dimensional 
regularization in mind, but is regularization scheme independent. We will therefore cross check our 
calculation in Sec.~\ref{Sec-4:EMT-PV} using the Pauli-Villars method, rederive the dimensional 
regularization result in Eq.~(\ref{Eq:EMT-tot-dim-result}), and confirm in this way that $N=4$ 
must be kept fixed in Eq.~(\ref{Eq:EMT-improve}) in dimensional regularization.

\newpage

\section{Pauli Villars Regularization}
\label{Sec-4:EMT-PV}

In the Pauli-Villars method, fictional particles are introduced with masses $\Lambda_i \gg m$ 
in order to regulate the divergent integrals. In an effective low energy theory, the cutoffs 
$\Lambda_i$ have a physical meaning and ensure that the result of the calculation applies only 
to low energies below the cutoffs $\Lambda_i$. In a renormalizable theory it is possible to take 
the limit $\Lambda_i\to\infty$ after the renormalization of the parameters of the theory.

To implement the Pauli-Villars regularization, we follow Callan, Coleman \& Jackiw \cite{Callan:1970ze}.
In $\Phi^4$ theory, it is necessary to introduce two fictional fields $\Phi_1$ and $\Phi_2$ in the 
Lagrangian as follows
\ba
    \mathcal{L}&=&\frac{1}{2}\bigl((\partial_{\mu}\Phi)^2-m^2\Phi\bigr)
    -\frac{1}{2}\bigl((\partial_{\mu}\Phi_1)^2-\Lambda_1^2\Phi_1^2\bigr)
    -\frac{1}{2}\bigl((\partial_{\mu}\Phi_2)^2-\Lambda_2^2\Phi_2^2\bigr) \nonumber \\
    && -\frac{1}{2}\delta_m\bigl(\Phi+a_1\Phi_1+a_2\Phi_2\bigr)^2-\frac{\lambda}{4!}\bigl(\Phi+a_1\Phi_1+a_2\Phi_2\bigr)^4
    \label{Eq:L-PV}
\ea
which makes the theory non-hermitian. The propagator of the unregularized theory
\begin{align*}
    \Pi(k^2)=\frac{1}{k^2-m^2+i\epsilon}
\end{align*}
is then modified to a propagator of the form
\begin{align*}
    \Pi_{\text{mod}}(k^2)=\frac{1}{k^2-m^2+i\epsilon}-\frac{a_1^2}{k^2-\Lambda_1^2+i\epsilon}-\frac{a_2^2}{k^2-\Lambda_2^2+i\epsilon}
    \, , \quad 
    a_1^2=\frac{m^2-\Lambda_2^2}{\Lambda_1^2-\Lambda_2^2} \, , \quad
    a_2^2=\frac{\Lambda_1^2-m^2}{\Lambda_1^2-\Lambda_2^2} \,.
 \end{align*}
Using the same on-shell renormalization scheme specified in Eq.~\eqref{onshell1},
the 1PI insertion in \eqref{onshell2} now becomes
\begin{equation}
    -iM^2(p^2)=-i \lambda \;\frac{1}{2} \;\int \frac{d^d k}{(2 \pi)^d}\Pi_{\text{mod}}(k^2)-i\delta_m. 
   \end{equation}
Thus, the mass counterterm in Eq.~(\ref{Eq:L-PV}) is of the form
\begin{equation}
    \delta_m=-\frac{\lambda}{2}\int \frac{d^4k_E}{(2\pi)^4}\biggl(\frac{1}{k_E^2+m^2}-\frac{m^2-\Lambda_2^2}{(\Lambda_1^2-\Lambda_2^2)(k_E^2+\Lambda_1^2)}-\frac{\Lambda_1^2-m^2}{(\Lambda_1^2-\Lambda_2^2)(k_E^2+\Lambda_2^2)}\biggr).
\end{equation}
The canonical EMT and the improvement term in the Pauli-Villars regularization are given by \cite{Callan:1970ze}
\ba
    T^{\mu\nu}_{\rm can,PV}
    &=& \partial^{\mu}\Phi\partial^{\nu}\Phi
        -\partial^{\mu}\Phi_1\partial^{\nu}\Phi_1
        -\partial^{\mu}\Phi_2\partial^{\nu}\Phi_2-g^{\mu\nu}\mathcal{L} \nonumber\\
    \label{Eq:EMT-improve-PV}
    \Theta^{\mu\nu}_{\rm imp,PV} 
    &=&
    -h\bigl(\partial^\mu\partial^\nu-g^{\mu\nu}\square\bigr)\bigl(\Phi^2-\Phi_1^2-\Phi_2^2\bigr)
\ea
with $h$ as defined in Eq.~(\ref{Eq:EMT-improve}).

To simplify our work, we note that
\ba
&&  -\frac{\lambda}{4!}\bra{\Omega}T\biggl(a_{p'}\bigl( \Phi^4+6c_1^2\phi_1^2\Phi^2+6c_2^2\phi_2^2\Phi^2\bigr)a_p^{\dagger}\biggr)\ket{\Omega}
    \nonumber\\
&&  \quad 
    =
    -\frac{\lambda}{2}\int \frac{d^4k_E}{(2\pi)^4}\biggl(\frac{1}{k_E^2+m^2}-\frac{m^2-\Lambda_2^2}{(\Lambda_1^2-\Lambda_2^2)(k_E^2+\Lambda_1^2)}-\frac{\Lambda_1^2-m^2}{(m_1^2-\Lambda_2^2)(k_E^2+\Lambda_2^2)}\biggr).
\ea
Hence, to order $\lambda$, we find 
\begin{gather*}
    -\frac{\lambda}{4!}\bra{\Omega}T\biggl(a_{p'} (\Phi^4+6c_1^2\Phi_1^2\Phi^2+6c_2^2\Phi_2^2\Phi^2 )a_p^{\dagger}\biggr)\ket{\Omega}-\frac{1}{2}\delta_m\bra{\Omega}T\biggl(a_{p'} (\Phi+a_1\Phi_1+a_2\Phi_2)^2a_p^{\dagger}\biggr)\ket{\Omega}=0
\end{gather*}
and our calculation reduces to evaluating the following expression for the total EMT
\begin{gather}
    \bra{p'}T^{\mu\nu}_{PV}\ket{p}  =F^{\mu\nu}(\Delta)+G^{\mu\nu}(\Delta)+H^{\mu\nu}(\Delta)+I^{\mu\nu}(\Delta)
\end{gather}
where,
\begin{align}
&F^{\mu \nu}(\Delta)=\bra{p'}(\partial^{\mu}\Phi\partial^{\nu}\Phi-\partial^{\mu}\Phi_1\partial^{\nu}\Phi_1-\partial^{\mu}\Phi_2\partial^{\nu}\Phi_2)\ket{p}
\nonumber\\
&G^{\mu\nu}(\Delta)=-\frac{g^{\mu\nu}}{2}\bra{p'}((\partial_{\rho}\Phi)^2-(\partial_{\rho}\Phi_1)^2-(\partial_{\rho}\Phi_2)^2)\ket{p}
\nonumber\\
&H^{\mu\nu}(\Delta)=-\frac{g^{\mu\nu}}{2}\bra{p'}(-m^2\Phi^2+\Lambda_1^2\Phi_1^2+\Lambda_2^2\Phi_2^2)\ket{p}
\nonumber\\
&I^{\mu\nu}(\Delta)=-h\bra{p'}(\partial^{\mu}\partial^{\nu}-g^{\mu\nu}\Box)(\Phi^2-\Phi_1^2-\Phi_2^2)\ket{p}.
\end{align}
We find, after letting $\Lambda_1\to \Lambda$ and $\Lambda_2 \to \Lambda$ with $\Lambda^2\gg \overline{m}^2$, that 
\begin{align}
&F^{\mu \nu}(\Delta)=\biggl[p'^{\mu}p^{\nu}+p'^{\nu}p^{\mu}+\int_0^1 dx\biggl(\frac{\lambda g^{\mu\nu}}{2(4 \pi)^2}\overline{m}^2\log{\frac{\overline{m}^2}{\Lambda^2}}+\frac{\lambda\Delta^{\mu}\Delta^{\nu}x(1-x)}{(4 \pi)^2}\log{\frac{\Lambda^2}{\overline{m}^2}}\biggr)\biggr]\nonumber\\
&G^{\mu\nu}(\Delta)=-g^{\mu\nu}\biggl[p'\cdot p+\int_0^1 dx\biggl(\frac{\lambda }{(4 \pi)^2}\overline{m}^2\log{\frac{\overline{m}^2}{\Lambda^2}}
    +\frac{\lambda\Delta^{2}x(1-x)}{2(4 \pi)^2}\log{\frac{\Lambda^2}{\overline{m}^2}}\biggr)\biggr]\nonumber\\
&H^{\mu\nu}(\Delta)=g^{\mu\nu}\biggl[m^2-\int_0^1 dx\frac{\lambda}{2(4 \pi)^2}m^2\log{\frac{\Lambda^2}{\overline{m}^2}}\biggr]\nonumber\\
&I^{\mu\nu}(\Delta)=h\biggl[2-\frac{\lambda}{(4 \pi)^2}\log{\frac{\Lambda^2}{\overline{m}^2}}\biggr](\Delta^{\mu}\Delta^{\nu}-g^{\mu\nu}\Delta^2)\label{PV}
\end{align}
with the details of these derivations given in App.~\ref{App:D} .

After combining the results Eq.~\eqref{PV}, simplifying, and letting $\Lambda\to \infty$, we arrive at the one-loop matrix elements for the total EMT,
\begin{align}\label{Eq:PV-final-result}
\bra{p'}T^{\mu\nu}_{\rm PV}\ket{p}=\frac{P^{\mu}P^\nu}{2}&-\frac{1}{3}\frac{\Delta^{\mu}\Delta^\nu-g^{\mu\nu}\Delta^2}{2}\nonumber\\
\qquad\qquad&+(\Delta^{\mu}\Delta^\nu-g^{\mu\nu}\Delta^2)\frac{\lambda}{(4\pi)^2}\int_0^1 dx\biggl(-(1-x)x+\frac{1}{6}\biggr)\log{(1-\frac{\Delta^2}{m^2}x(1-x))}.
\end{align}

From Eq.~(\ref{Eq:PV-final-result}) we read of the results for  $A(t)_{\rm one\!-\!loop}$ 
and $D(t)_{\rm one\!-\!loop}$ in complete agreement with \eqref{Eq:EMT-tot-dim-result}.

\newpage

\section{\boldmath Properties of $D(t)$}
\label{Sec-5:properties-of-D(t)}

After computing the EMT form factors in the $\Phi^4$ theory in one-loop order in dimensional
regularization in Eq.~(\ref{Eq:EMT-tot-dim-result}) and double-checking the results in Pauli-Villars
regularization, we now turn to a discussion of their physical properties. In the following, the
``one-loop'' subscript will be dropped for notational simplicity.

The one-loop calculation shows that the EMT form factor $A(t) = 1$ remains a trivial constant, while 
$D(t)$ exhibits an non-trivial $t$-dependence which makes it worth discussing. For convenience, let us 
quote again the result  
\be
\label{Eq:D-final}
    D(t) = -\frac{1}{3}
    +\frac{2\lambda}{(4\pi)^2}\int_0^1 dx\biggl(\frac{1}{6}-(1-x)x\biggr)
    \log\biggr(1-\frac{t}{m^2}x(1-x)\biggl) \,.
\ee
The value of the $D$-term, the form factor $D(t)$ at $t=0$ is $\lambda$-independent 
and given by 
\be
\label{Eq:D-term}
    D = -\frac{1}{3} \,.
\ee

The integral in Eq.~(\ref{Eq:D-final}) can be solved in closed form for $t\neq0$. Below the threshold for particle 
production $t<4m^2$, the form factor $D(t)$ is real. Above $t>4m^2$, it develops an imaginary part. 
The expressions are given by
\begin{alignat}{5}
    {\rm Re\,}D(t) = && -\frac13 + \frac{\lambda}{24\pi^2}
    & \times
    \begin{cases}  \displaystyle
    \biggl[\frac{2m^2}{t}-\frac{1}{6}-\frac{2m^2}{t}\sqrt{1-\frac{4m^2}{t}}\;{\rm arsinh}\biggl(\frac{\sqrt{-t}}{2m}\biggr)\biggr]      & \mbox{for} \quad \; t < 0,  \\
    { } \\ \displaystyle
    \biggl[\frac{2m^2}{t}-\frac{1}{6}-\frac{2m^2}{t}\sqrt{\frac{4m^2}{t}-1}\;{\rm arcsin }\biggl(\frac{\,\sqrt{ t}}{2m}\,\biggr)\biggr] & \mbox{for} \quad 0 < t < 4m^2, \\
    { } \\ \displaystyle
    \biggl[\frac{2m^2}{t}-\frac{1}{6}-\frac{2m^2}{t}\sqrt{1-\frac{4m^2}{t}}\,{\rm arcosh}\biggl(\frac{\sqrt{t}}{2m}\biggr)\biggr]       & \mbox{for} \quad t > 4m^2, 
    \end{cases} \nonumber\\
    { } \nonumber\\
    {\rm Im\,}D(t) = && \frac{\lambda}{24\pi}
    & \times
    \begin{cases}  
    \displaystyle 0\, \hspace{67mm} & \mbox{for} \quad  t\le 4m^2 , \\
    { } \\
    \displaystyle \;\frac{m^2}{t}\sqrt{1-\frac{4m^2}{t}}  & \mbox{for} \quad t > 4m^2 .
    \end{cases}
    \label{Eq:Dabove+below}
\end{alignat}
In the vicinity of $t=0$, the form factor is continuous and (infinitely) differentiable, with a unique
Taylor expansion 
\be\label{Eq:D-Taylor}
    D(t) = -\frac{1}{3} + \frac{\lambda}{1440\pi^2} \biggl[
     \frac{t}{m^2}
    +\frac{t^2}{7\, m^4}
    +\frac{t^3}{42\,m^6}
    +\frac{t^4}{231\, m^8}
    + \dots \biggr] \;,
\ee
which is independent of whether one approaches the point $t=0$ from the region of positive or 
negative $t$. The limits $t\to \pm\infty$ in the respective terms in Eq.~(\ref{Eq:Dabove+below})
exist and are given by 
\be\label{Eq:D-inf}
    D_{\inf} =    
    \lim_{t\to-\infty}D(t) =    
    \lim_{t\to\infty}{\rm Re\,}D(t) = -\frac{1}{3} - \frac{\lambda}{144\pi^2},  \quad
    \lim_{t\to\infty}{\rm Im\,}D(t) = 0\,.
\ee
At the threshold $t=4m^2$, the real part of $D(t)$ is continuous but not differentiable and 
assumes at the cusp the value
\be\label{Eq:D-cusp}
    D_{\rm cusp} = {\rm Re}\,D(4m^2) = -\frac{1}{3} + \frac{\lambda}{72\pi^2}\,.
\ee
The imaginary part of $D(t)$ is non-zero only above the threshold $t=4m^2$ 
and is always non-negative. It exhibits a global maximum at $t=6m^2$ with the value 
\be\label{Eq:ImDmax}
    {\rm Im}\,D_{\max} = {\rm Im}\,D(6m^2) = \frac{\lambda}{144\sqrt{3}\pi}\,.
\ee

\begin{figure}[t!]
\begin{centering}
\includegraphics[width=4.2cm]{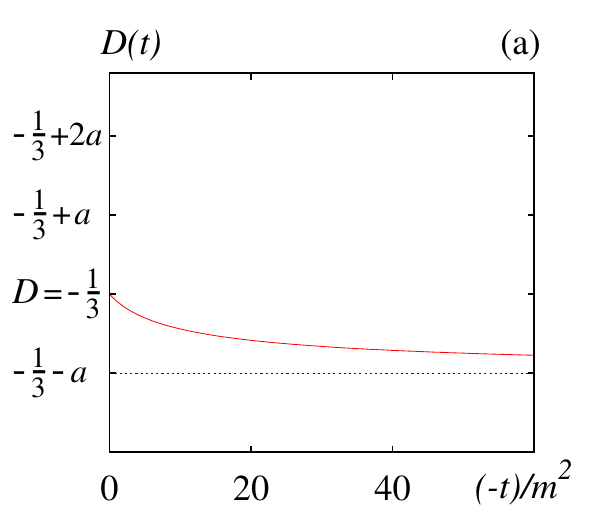} \
\includegraphics[width=4.2cm]{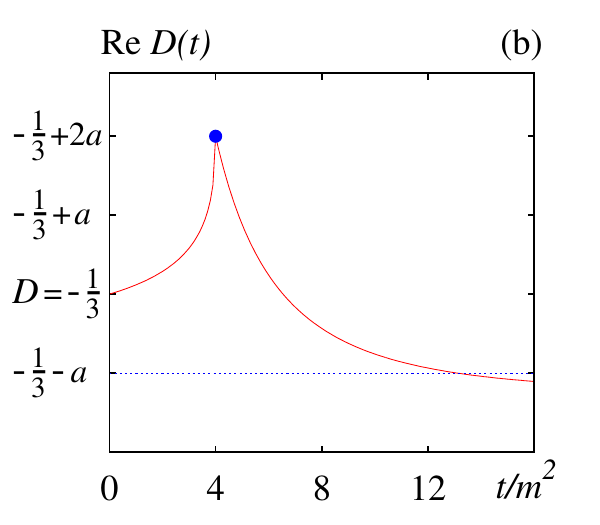} \
\includegraphics[width=4.2cm]{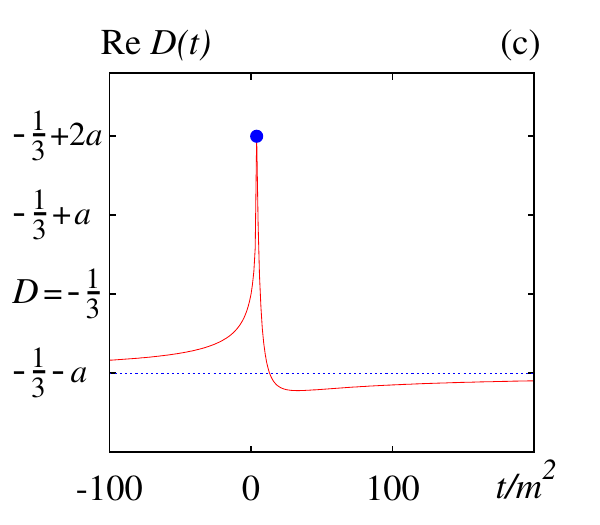 } \
\includegraphics[width=4.2cm]{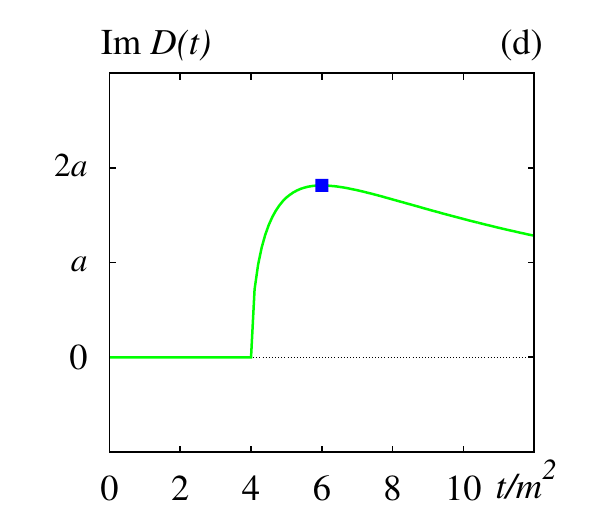} 
\par\end{centering}
\caption{\label{Fig:D-of-t} The form factor $D(t)$ 
    in $\Phi^4$ theory at one loop as function of $t$ in units of $m^2$. 
    In all figures, 1 unit on the $y$-axis corresponds to $a=\lambda/(144\pi^2)$.
    (a) The space-like region plotted as function of $(-t)>0$. 
    (b) ${\rm Re}\,D(t)$ in the time-like region.
    (c) The same as (a) and (b) but showing a wide $t$-range. 
    (d) ${\rm Im}\,D(t)$ which is non-zero only above the threshold $t=4m^2$.
    The vertical lines in (a)--(c) are the asymptotic values $D_{\inf}$ 
    defined in Eq.~(\ref{Eq:D-inf}). 
    The dots in (b) and (c) mark the cusp defined in Eq.~(\ref{Eq:D-cusp}).
    The square in (d) denotes the maximum of ${\rm Im}\,D(t)$ in Eq.~(\ref{Eq:ImDmax}). }
\end{figure}

In Fig.~\ref{Fig:D-of-t} we show the real and imaginary parts of $D(t)$
as functions of $t$ in units of $m^2$ under the assumption that $\lambda$ is small and 
the one-loop approximation is justified. 
For $\lambda\ll 1$, it is always ${\rm Re}\,D(t)<0$ while ${\rm Im}\,D(t)\ge 0$. 
If we literally assume $\lambda$ to be infinitesimally small, then $D(t)\simeq -\frac13$ is a 
trivial constant. In order to show a non-trivial $t$ dependence 
of $D(t)$, we choose the unit on the $y$-axis in Fig.~\ref{Fig:D-of-t} to correspond
to $a=\lambda/(144\pi^2)$.

In Fig.~\ref{Fig:D-of-t}a, we show the space-like region where $D(t)$ is real and
plotted as a function of $(-t)>0$ as it is customary for form factors measured in 
scattering experiments. We see that $D(t)$ decreases monotonically from $D=-\frac13$
to the asymptotic value $D_{\inf}$ in Eq.~(\ref{Eq:D-inf}), i.e.\ the absolute value
$|D(t)|$ increases which we shall comment on below.
The Fig.~\ref{Fig:D-of-t}b shows ${\rm Re}\,D(t)$ in the time-like region around the 
threshold $t=4m^2$ above which $D(t)$ becomes complex. The dot at $t=4m^2$ marks the 
cusp defined in Eq.~(\ref{Eq:D-cusp}). 
In Fig.~\ref{Fig:D-of-t}c, we plot a wide $t$-range to illustrate how 
${\rm Re}\,D(t)$ approaches logarithmically the asymptotic value $D_{\inf}$ in 
Eq.~(\ref{Eq:D-inf}).
The Fig.~\ref{Fig:D-of-t}d shows ${\rm Im}\,D(t)$, which is non-zero only above the 
threshold and exhibits a maximum at $t=6m^2$, which is defined in Eq.~(\ref{Eq:ImDmax}) 
and marked in Fig.~\ref{Fig:D-of-t}d by a square. ${\rm Im}\,D(t)$ approaches asymptotically 
zero for $t\to\infty$.
Higher loops would generate further cusps and thresholds in respectively ${\rm Re}\,D(t)$
and ${\rm Im}\,D(t)$ at $t=4n^2m^2$ for integer $n>1$. However, the relative size of the cusps 
(relative to $D=-\frac13$) or thresholds would be proportional to $\lambda^n$ and not visible 
on the scale of Fig.~\ref{Fig:D-of-t} for $\lambda\ll 1$. 

The growth of $|D(t)|$ in $\Phi^4$-theory for $t\to-\infty$ is interesting.
Indeed, hadronic models
\cite{Ji:1997gm,Goeke:2007fp,Goeke:2007fq,Wakamatsu:2007uc,Cebulla:2007ei,Abidin:2008ku,Broniowski:2008hx,Frederico:2009fk,Kim:2012ts,Jung:2014jja,Jung:2013bya,Chakrabarti:2015lba,Perevalova:2016dln,Kumar:2017dbf,Freese:2019bhb,Neubelt:2019sou,Mamo:2019mka,Sun:2020wfo,Kim:2020nug,Kim:2020lrs,Krutov:2020ewr,Chakrabarti:2020kdc,Fu:2022rkn,Mamo:2022eui,Won:2022cyy,GarciaMartin-Caro:2023klo,Amor-Quiroz:2023rke,Cao:2023ohj,Xu:2024cfa},
lattice QCD 
\cite{Hagler:2003jd,Gockeler:2003jfa,LHPC:2007blg,Shanahan:2018pib,Alexandrou:2019ali,Pefkou:2021fni,Hackett:2023nkr,Hackett:2023rif,Delmar:2024vxn},
perturbative QCD 
\cite{Tong:2022zax,Guo:2023qgu,Tong:2021ctu},
data-based determinations 
\cite{Pasquini:2014vua,Kumano:2017lhr,Burkert:2018bqq},
and QCD sum rules 
\cite{Azizi:2019ytx,Ozdem:2020ieh,Azizi:2020jog,Aliev:2020aih}
yield a monotonically decreasing $|D(t)|$ for increasing $(-t)$
which is a behavior shared by all hadronic form factors.\footnote{Some 
    form factors like the neutron electric form factor $G_E^n(t)$ or the nucleon gravitomagnetic 
    form factor $B(t)$ vanish at $t=0$ and increase initially with $(-t)$. But eventually 
    their magnitudes also decrease like other hadronic form factors.}
This unfamiliar behavior of $D(t)$, from a hadronic physics perspective,
deserves some comments.

The $\Phi^4$ interaction for $\lambda>0$ is repulsive \cite{Beg:1984yh} in contrast to the strong 
interaction, which is attractive and forms bound states. This raises the question whether the 
growth of $|D(t)|$ might be a feature of a repulsive interaction. 
In fact, if we could continue the theory (\ref{Eq:L}) to $\lambda<0$, then the
interaction would become attractive and our $|D(t)|$ would exhibit a ``more hadronic behavior'' 
and decrease for $(-t)\to\infty$. The continuation to negative $\lambda$ would not affect our
perturbative calculation, but it would make the theory (\ref{Eq:L}) unbound from below and 
the vacuum unstable.

However, in this context it is interesting to note that the (complex) $|\Phi|^4$ theory continued 
analytically to $\lambda<0$ with an additional  positive term $C\,|\Phi|^6$ in the Lagrangian is 
a well-defined (albeit effective and non-renormalizable) theory. This theory admits 
non-perturbative soliton solutions referred to as $Q$-balls \cite{Coleman:1985ki} which exhibit 
EMT properties familiar from nuclear and hadronic physics \cite{Mai:2012yc,Mai:2012cx}. 
In this theory it is possible to take a certain (so-called $Q$-cloud) limit
\cite{Alford:1987vs}, where the additional term  $C\,|\Phi|^6$ in the Lagrangian becomes irrelevant 
and can effectively be set to zero, i.e.\ one basically deals with a complex $|\Phi|^4$ theory
continued to negative $\lambda$. Remarkably, the appropriately rescaled $Q$-cloud solutions 
encountered in this limit still exhibit hadron-type EMT properties \cite{Cantara:2015sna}.

These considerations illustrate that the growth of $|D(t)|$ in $\Phi^4$-theory for 
$t\to-\infty$ in our calculation is natural. 
(Notice that $Q$-balls and $Q$-clouds exist only in complex scalar theories, while our
calculation was in real $\Phi^4$ theory. But our 1-loop calculation would look basically 
the same in complex $\Phi^4$ theory.)

Of course, one may also pragmatically argue that in the weak coupling limit needed to 
justify the perturbative treatment, $D(t)\simeq -\frac13$ is nearly constant as one 
would expect for a form factor of a particle with no internal structure. 
It is also not surprizing that in $\Phi^4$ theory the form factors (trivially for $A(t)$ 
and non-trivially for $D(t)$) approach asymptotically a constant for $(-t)\to\infty$ and
do not vanish like hadronic form factors do. The latter reflects the fact that hadrons are
composed particles. It is unlikely for a hadron to stay intact after absorbing a large 
momentum transfer. The situation is of course distinctly different in $\Phi^4$ theory.

\newpage
\section{\boldmath Dispersion relation for $D(t)$ in $\Phi^4$ theory}
\label{Sec-6:dispersion-relation}

The form factor $D(t)$ satisfies a dispersion relation \cite{Pasquini:2014vua}, see also
\cite{Kumano:2017lhr}, which provides a powerful test for the calculation. 
In our case, the form factor does not go to zero for $|t|\to\infty$. In this case, one can either
formulate an unsubtracted dispersion relation for $D(t) - D_{\inf}$ which does go to zero for $|t|\to\infty$,
or subtracted dispersion relation for $D(t) - D(0)$ which does not go to zero for $|t|\to\infty$.
Both variants are given by
\ba\label{Eq:disp-unsub}
    D(t) - D_{\inf} &=& \frac1\pi \int_{4m^2}^\infty ds\; \frac{{\rm Im}\,D(s)}{s-t}  
    \quad \; \mbox{unsubtracted case for} \, \quad t < 4 m^2 \, , \\ \nonumber\\
    \label{Eq:disp-sub} 
    D(t) - D(0)     &=& \frac1\pi \int_{4m^2}^\infty ds\;\frac{{\rm Im}\,t\,D(s)}{s(s-t)}  
    \quad \mbox{subtracted case for} \quad t < 4 m^2 \, .
\ea
Notice that ${\rm Im}\,D_{\inf}={\rm Im}\,D(0)=0$ was used to simplify the expressions on the
right-hand side. In either case, one must subtract the tree-level term $D(0)$, which is of order 
$\lambda^0$. Only the loop contribution of order $\lambda$ has non-trivial analytical properties. 
We checked analytically and numerically that our results (\ref{Eq:D-final},~\ref{Eq:Dabove+below}) 
satisfy the dispersion relations in Eq.~(\ref{Eq:disp-unsub},~\ref{Eq:disp-sub}).
This is an important cross check for the calculation.

The form factor $A(t)=1$ satisfies the dispersion relation trivially. Also, here it is necessary 
to formulate the dispersion relation for the difference $A(t)-1$, which is the quantity that goes 
to zero for $|t|\to\infty$. The real and imaginary parts of $A(t)-1$ vanish, and the 
unsubtracted dispersion becomes zero equals zero.

\section{Radii}
\label{Sec-7:radii}

One of the reasons for the interest in EMT form factors is related to their attractive interpretations.
Fourier transforms of the EMT form factors give insights into the distribution of matter
and internal forces inside an extended particle. For particles whose radius is much larger than 
the Compton wavelength, one can discuss 3D densities \cite{Polyakov:2002yz,Polyakov:2018zvc}.
For particles of any mass, including massless particles, one can discuss 2D densities 
\cite{Lorce:2018egm,Freese:2021czn}.
It is interesting to investigate aspects of this interpretation in the $\Phi^4$ theory,
and study the mean square radii of several quantities.

Let us begin with what has been referred to as the mean square radius of the matter
distribution in some works \cite{Xu:2024cfa}. This radius is defined in terms of the 
derivative of the EMT form factor $A(t)$ at $t=0$ as follows
\be\label{Eq:r2-matter}
    \langle r^2_{\rm mat}\rangle  = - 6\,A'(0)\,.
\ee
Since we found $A(t)=1$, we see that the mean square radius of the matter distribution 
in the $\Phi^4$ theory is 
\be
    \langle r^2_{\rm mat}\rangle  = 0\,.
\ee
In the 2D light-front formulation, the slope of $A(t)$ at $t=0$ (with a prefactor of minus $4$)
defines the so-called light-front momentum mean square radius \cite{Freese:2021czn}, which is
of course also zero.

Another interesting quantity is the mean square scalar radius associated with the trace of the EMT. 
The form factor of this trace of the EMT can be expressed in terms of $A(t)$ and $D(t)$ based 
on Eq.~(\ref{Eq:def-EMT-FF}). The mean square radius of the EMT trace operator \cite{Goeke:2007fp} 
is defined as  
\be
       \langle r^2_{\rm tr}\rangle  =   
       \langle r^2_{\rm mat}\rangle  -
       \frac{3}{2m^2}\;(1+3D)\,.
\ee
This quantity has been referred to in literature also as the mass radius \cite{Kharzeev:2021qkd} 
or scalar radius \cite{Xu:2024cfa}. With our results for the EMT form factors, we obtain
\be
       \langle r^2_{\rm tr}\rangle  =  0\,.
\ee

Finally, let us discuss the mechanical radius, whose definition does not involve derivatives
of form factors, but rather is given by  \cite{Polyakov:2018guq,Polyakov:2018zvc}
\be
    \langle r^2_{\rm mech}\rangle = \frac{6\,D(0)}{\int_{-\infty}^0 dt\;D(t)}\;.
\ee
In the $\Phi^4$ theory, the numerator is finite while the integral in the denominator 
diverges such that one obtains 
\be
    \langle r^2_{\rm mech}\rangle = 0\;.
\ee

The mean square radii discussed in these examples vanish in the $\Phi^4$ theory
at one-loop order, which shows that the described particle is point-like.

\newpage
\section{Conclusions}
\label{Sec-8:conclusions}

In this work, a study of EMT form factors was presented in $\Phi^4$ theory in the 
weak-coupling regime at one-loop. This study is of interest as the EMT form
factors of hadrons have attracted a lot of attention in the recent past
\cite{Polyakov:2018zvc,Burkert:2023wzr}. In view of the complexity of QCD,
EMT studies in simpler theories can be very instructive. After free field theory,
the next-to-simplest case is to solve an interacting theory perturbatively to 
lowest non-trivial order. 

In free spin-zero theory, the EMT form factors can be evaluated exactly with  
$A(t)_{\rm free}=1$ and $D(t)_{\rm free}=-1$ \cite{Pagels:1966zza}.
This implies for the $D$-term $\Dfree=-1$  of a free spin-zero boson
\cite{Hudson:2017xug}. In the weak coupling regime $\lambda\ll 1$,  
the cross sections in $\Phi^4$ theory is proportional to $\lambda^2$, 
i.e.\ the $\Phi$ quanta are nearly free particles. One could therefore 
expect the $D$-term in a weakly interacting $\Phi^4$ theory to be very 
similar to the free theory. But this is not the case.

In this work, we have shown that the inclusion of a $\Phi^4$ interaction reduces the 
$D$-term from $\Dfree=-1$ in free case to $D = -\frac13$ no matter how small the coupling 
constant $\lambda$ is. This was speculated without proof in Ref.~\cite{Hudson:2017xug} 
based on the necessity to introduce an improvement in the EMT in scalar field theory 
which is needed to render EMT matrix elements finite in the presence of quantum interactions 
\cite{Callan:1970ze}. This result is supported by calculations in another renormalizable scalar 
theory, the $\Phi^3$ theory \cite{Dotson:new}, and in line with observations in the literature 
that the $D$-term is the particle property which is most sensitive to the dynamics in a system  
\cite{Polyakov:2018zvc,Burkert:2023wzr,Hudson:2017xug,Hudson:2017oul}.

The calculation was carried out at one-loop in dimensional regularization 
and checked independently in Pauli-Villars regularization. The consistency of the 
calculations was checked in several ways: the EMT is of course conserved also at one-loop, 
the form factor $A(t)$ satisfies the constraint $A(0)=1$ imposed by Lorentz symmetry, 
and both EMT form factors $A(t)$  and $D(t)$ are renormalization scale independent. 

Assuming that $\lambda\ll 1$ and ${\cal O}(\lambda^2)$ can be safely neglected, our results yield 
the ``full results'' for the EMT form factors in $\Phi^4$ theory in the weak coupling regime.
The form factor $A(t)$ is not altered by quantum corrections and remains $A(t)=1$ also in the 
interacting case. For $D(t)$ the situation is different: at $t=0$ it assumes the exact value 
$D=-\frac13$ which is a drastic change from $\Dfree = -1$, while for $t\neq 0$ it exhibits a 
non-trivial $t$-dependence.

In the space-like region $t<0$ the form factor $D(t)$ shows an unfamiliar behavior from the
point of view of hadronic physics with $|D(t)|$ slowly growing as $(-t)$ becomes large,
while the form factors of hadrons decrease. This behavior can be understood considering 
that the quanta in $\Phi^4$ theory are elementary fields, not composed particles. 
The weak increase of $|D(t)|$ which logarithmically approaches a constant for asymptotically 
large $(-t)$ might be related to the fact that the interaction in $\Phi^4$ theory is repulsive. 
This interesting point deserves further studies.

Exploring the field theoretical setting, we continued analytically our diagrammatic
calculation to time-like $t>0$. At the threshold for two-particle 
production, $t=4m^2$ nothing unusual happens to $A(t)$ which remains equal to unity 
also in the entire time-like region. The form factor $D(t)$ becomes complex for $t>4m^2$.
We explicitly determined the analytical expressions for ${\rm Re}\,D(t)$ and
${\rm Im}\,D(t)$ and showed that they satisfy the dispersion relation \cite{Pasquini:2014vua}.

Our study was rounded up by an investigation of several mean square radii 
introduced in recent literature.  
In particular, we investigated the mean square radius of the matter distribution
$\langle r^2_{\rm mat}\rangle  = - 6\,A'(0)$,
the mean square radius of the EMT trace 
$\langle r^2_{\rm tr}\rangle  =  - 6\,A'(0)  - \frac{3}{2m^2}\;(1+3D)$, 
and the mechanical radius 
$\langle r^2_{\rm mech}\rangle = 6\,D/ \int_{-\infty}^0 dt\;D(t)$.
All these radii vanish in the $\Phi^4$ theory. 
On the one hand, this is not unexpected since the particles in the 
$\Phi^4$ theory are elementary and point-like. 
On the other hand, it is reassuring to observe that the various mean square radii 
recently introduced to describe EMT properties of hadrons yield zero for
pointlike particles and hence have a sound physical basis. In this sense, our study 
contributes to a better understanding of EMT properties and their interpretation.

We stress that the $\phi^4$ results are not representative for ``all weakly coupled theories'' 
and ``all elementary particles''.  QED is a well-known counterexample
\cite{Berends:1975ah,Milton:1976jr,Kubis:1999db,Donoghue:2001qc,Varma:2020crx,Metz:2021lqv,Freese:2022jlu}.
The long-range nature of the Coulomb force \cite{Donoghue:2001qc,Varma:2020crx},
based on the masslessness of the photon in gauge theory, gives rise 
to very different EMT properties for charged particles independently of their spin. 
For instance, in one-loop order in QED the EMT form factors of the electron vanish for 
asymptotically large $(-t)$ and no mean square radii can be defined \cite{Freese:2022jlu}. 

While our results are sufficient in the weak coupling limit, it would be 
nevertheless instructive to study EMT properties in $\Phi^4$ theory beyond one-loop.
The interesting feature is the running of the coupling constant $\lambda\to\lambda(\mu)$
starting at two-loop order. Form factors of conserved external currents are expected 
to be renormalization scale independent. It would be interesting to see how this is realized
in a two-loop calculation in $\Phi^4$ theory. This question goes beyond the scope 
of this work and is left to future studies.

\ 

\noindent
{\bf Acknowledgments.} 
Discussions with John Collins, Andrew Dotson, Gerald Dunne, Luchang Jin, Barbara Pasquini, 
and Matthew Sievert played important roles to motivate and carry out this work. 
The author is indebted to Andrew Dotson for checking parts of the calculation and 
Peter Schweitzer for help with  manuscript preparation. 
This work was supported by National Science Foundation under Award No.\ 2111490, and 
U.S.\ Department of Energy under the umbrella of the Quark-Gluon Tomography (QGT) 
Topical Collaboration with Award No. DE-SC0023646.

\appendix

\section{Dimensional regularization}

This Appendix contains for reference the details of the EMT calculation in dimensional regularization.

\subsection{\boldmath The expressions for $f^{\mu\nu}(\Delta)$, $g(\Delta)$, $h(\Delta)$, $j(\Delta)$}
\label{App:A}

Let us first derive the expressions  
for $f^{\mu\nu}(\Delta)$, $g(\Delta)$, $h(\Delta)$, $j(\Delta)$
introduced in Eq.~(\ref{Eq:EMT-can-2}) which emerge from respectively 
evaluating matrix elements of the operators 
$\partial^{\mu}\Phi\partial^{\nu}\Phi$,
$\partial_\rho\Phi\partial^\rho\Phi$, 
$\Phi^2$,
$\Phi^4$ in Eq.~(\ref{Eq:EMT-can-1}).
We find
\begin{align}   \label{eq:7}
\setlength{\jot}{10pt}
    & \hspace{-2cm}
    \bra{\Omega}T(a_{p'}\partial^{\mu}\Phi(x)\partial^{\nu}\Phi(x)a_{p}^{\dagger}e^{-i\int \frac{\lambda}{4!}\Phi^4d^4y})\ket{\Omega}
    \nonumber\ph\\
    &
    =(p'^{\mu}p^{\nu}+p'^{\nu}p^{\mu})e^{i\Delta\cdot x}-\frac{i\lambda}{4!}\int d^4y\bra{0}T(a_{p'}\partial^{\mu}\Phi(x)\partial^{\nu}\Phi(x) a_{p}^{\dagger}\Phi^4(y))\ket{0}
     \nonumber\ph\\
    &=\biggl(p'^{\mu}p^{\nu}+p'^{\nu}p^{\mu}
    +i\lambda \int \frac{d^4k}{(2\pi)^4} \frac{k^{\mu} }{k^2-m^2+i\epsilon}
    \frac{(k-\Delta)^{\nu} }{(k-\Delta)^2-m^2+i\epsilon}
    \biggr)e^{i\Delta\cdot x} 
     \nonumber\ph\\
    &=(p'^{\mu}p^{\nu}+p'^{\nu}p^{\mu} +\lambda \,f^{\mu\nu}(\Delta))e^{i\Delta\cdot x},  \ph\\
\label{eq:8}
     &\hspace{-2cm}
     \frac{g^{\rho \sigma}}{2}\bra{\Omega}T(a_{p'}\partial_{\rho}\Phi(x)\partial_{\sigma}\Phi(x)a_{p}^{\dagger}e^{-i\int \frac{\lambda}{4!}\Phi^4d^4y})\ket{\Omega}\nonumber\ph\\
    &=\frac{g^{\rho \sigma}}{2}[p'_{\rho}p_{\sigma}e^{i\Delta \cdot x}+p'_{\sigma}p_{\rho}e^{i\Delta \cdot x}-\frac{i\lambda}{4!}\int d^4y\bra{0}T(a_{p'}\partial_{\rho}\Phi(x)\partial_{\sigma}\Phi(x) a_{p}^{\dagger}\Phi^4(y))\ket{0}]   \nonumber\ph\\
    &=\biggl(p'\cdot p
    +\frac{i\lambda}{2}\int\frac{d^4k}{(2\pi)^4}\frac{k_{\rho}}{k^2-m^2+i\epsilon}\frac{(k-\Delta)^{\rho}}{(k-\Delta)^2-m^2+i\epsilon}\biggr)e^{i\Delta \cdot x}   \nonumber\ph\\
    &=(p'\cdot p
    -\lambda g(\Delta))e^{i\Delta \cdot x} \ph\\
\label{eq:9}
    &\hspace{-2cm}
    \frac{(m^2+\delta_m)}{2} \bra{\Omega}T(a_{p'}\Phi(x)\Phi(x)a_{p}^{\dagger}e^{-i\int \frac{\lambda}{4!}\Phi^4d^4y})\ket{\Omega}   \nonumber\\
    &=\frac{(m^2+\delta_m)}{2}\biggl(2e^{i\Delta \cdot x}-\frac{i\lambda}{4!}\int d^4y\bra{0}T(a_{p'}\Phi(x)\Phi(x) a_{p}^{\dagger}\Phi^4(y))\ket{0}\biggr)   \nonumber\\
    &=(m^2+\delta_m)\biggl(1+\frac{i\lambda}{2} \int  \frac{d^4k}{(2\pi)^4} \frac{1}{k^2-m^2+i\epsilon}
    \frac{1}{(k-\Delta)^2-m^2+i\epsilon}\biggr)e^{i\Delta \cdot x}   \nonumber\\
    &=(m^2+\delta_m)(1+\lambda h(\Delta))e^{i\Delta \cdot x}, \ph\\
\label{eq:10}
    &\hspace{-2cm} \frac{\lambda}{4!}\bra{\Omega}T(a_{p'}\Phi(x)\Phi(x)\Phi(x)\Phi(x) a_{p}^{\dagger}e^{-i\int \frac{\lambda}{4!}\Phi^4d^4y})\ket{\Omega} \nonumber\\
    &=\frac{\lambda}{2}\int\frac{d^4k}{(2\pi)^4}\frac{ie^{i(p'-p)\cdot x}}{k^2-m^2+i\epsilon}  
    =\lambda j(\Delta)e^{i\Delta\cdot x}. \ph
 \end{align}
 Combining Eqs.~(\ref{eq:7}--\ref{eq:10}) and 
 letting $x \to 0$, we obtain the result for the canonical EMT quoted in Eq.~(\ref{Eq:EMT-can-1}). 

\newpage
\subsection{Evaluation of canonical EMT in dimensional regularization}
 \label{App:B}

We shall make use of the following integrals in dimensional regularization 
in Euclidean space-time where $k^0=ik^0_E$ 
\begin{align*}
    &\int \frac{dk^4}{(2\pi)^4}\frac{1}{(k^2-\Sigma)^n}\xrightarrow{ }  
    (-1)^ni\int\frac{d^4k_E}{(2\pi)^4}\frac{1}{(k_E^2+\Sigma)^n}
    =
    (-1)^ni\frac{1}{(4 \pi)^{d / 2}} \frac{\Gamma\left(n-\frac{d}{2}\right)}{\Gamma(n)}\left(\frac{1}{\Sigma}\right)^{n-\frac{d}{2}},\\
    &\int \frac{dk^4}{(2\pi)^4}\frac{k^2}{(k^2-\Sigma)^n}\xrightarrow{ } 
    (-1)^{n+1}i\int\frac{d^4k_E}{(2\pi)^4}\frac{k_E^2}{(k_E^2+\Sigma)^n}
    =(-1)^{n+1}i\frac{1}{(4 \pi)^{d / 2}}\frac{d}{2} \frac{\Gamma\left(n-\frac{d}{2}-1\right)}{\Gamma(n)}\left(\frac{1}{\Sigma}\right)^{n-\frac{d}{2}-1}.
\end{align*}
After changing the measure of our integrals from 4 to the arbitrary dimension $d$, the
integrals are evaluated as follows
 \begin{align*}
    f^{\mu\nu}(\Delta)
    &=i\mu^{4-d} \int \frac{d^d k}{(2\pi)^d} \frac{k^{\mu}}{k^2-m^2+i\epsilon}
    \frac{(k-\Delta)^{\nu}}{(k-\Delta)^2-m^2+i\epsilon}\\
    &=i\mu^{4-d} \int \frac{d^d k}{(2\pi)^d}k^{\mu}(k-\Delta)^{\nu}\int_0^1 dx
    \frac{1}{(k^2-m^2+i\epsilon+(\Delta^2-2k\cdot \Delta)x)^2}
    \quad\\
    &=i\mu^{4-d} \int \int_0^1 \frac{d^d k dx}{(2\pi)^d}(k^{\mu}k^{\nu}-\Delta^{\mu}\Delta^{\nu}x+\Delta^{\mu}\Delta^{\nu}x^2)
    \frac{1}{(k^2-\overline{m}^2(x,t)+i\epsilon)^2}\\
    &=-\mu^{4-d} \int_0^1 dx\biggl(-g^{\mu\nu}\frac{\Gamma(1-\frac{d}{2})}{2\overline{m}(x,t)^{2-d}(4\pi)^{\frac{d}{2}}}-
    \Delta^{\mu}\Delta^{\nu}(1-x)x\frac{\Gamma(2-\frac{d}{2})}{\overline{m}(x,t)^{4-d}(4\pi)^{\frac{d}{2}}}\biggr)\\
    g(\Delta)
    &=-i\frac{\mu^{4-d}}{2} \int \frac{d^d k}{(2\pi)^d} \frac{k_{\mu}}{k^2-m^2+i\epsilon}
    \frac{(k-\Delta)^{\mu}}{(k-\Delta)^2-m^2+i\epsilon}\\
    &=-i\frac{\mu^{4-d}}{2}\int \frac{d^d k}{(2\pi)^d}\int_0^1 dx \frac{(k+\Delta x)_{\mu}(k+\Delta(x-1))^{\mu}}{(k^2-\overline{m}^2(x,t)+i\epsilon)^2}\\
    &=-i\frac{\mu^{4-d}}{2}\biggl[\int \frac{d^dk}{(2\pi)^d}\int_0^1 dx \frac{k^2}{(k^2-\overline{m}^2(x,t)+i\epsilon)^2}-\int \frac{d^d k}{(2\pi)^d}\int_0^1 dx \frac{\Delta^2 (1-x)x}{(k^2-\overline{m}^2(x,t)+i\epsilon)^2}\biggr]\\
    &=\int_0^1 dx\frac{\mu^{4-d} }{2}\biggl(-\frac{\Gamma(1-\frac{d}{2})d}{2\overline{m}(x,t)^{2-d}(4\pi)^{\frac{d}{2}}}-\Delta^2(1-x)x\frac{\Gamma(2-\frac{d}{2})}{\overline{m}(x,t)^{4-d}(4\pi)^{\frac{d}{2}}}\biggr)\\
    h(\Delta)
    & =\frac{i \mu^{4-d}}{2} \int  \frac{d^d k}{(2\pi)^d}\int_0^1 dx \frac{1}{(k^2-\overline{m}^2(x,t)+i\epsilon)^2}
      =-\frac{\mu^{4-d}}{2(4\pi)^{\frac{d}{2}}}\int_0^1 dx \overline{m}(x,t)^{d-4}\Gamma(2-\frac{d}{2}) \\
    j(\Delta)
    & =\frac{\mu^{4-d} }{2}\int\frac{d^d k}{(2\pi)^d}\frac{i}{k^2-m^2+i\epsilon}
     =\frac{\mu^{4-d}}{2(4\pi)^\frac{d}{2}}\frac{\Gamma(1-\frac{d}{2})}{(m^2)^{(1-\frac{d}{2})}}.
\end{align*}
Using $d=4-\epsilon$ and combining the terms gives 
\begin{align*}
   &\lambda\left(f^{\mu\nu}(\Delta)+g^{\mu\nu}\left(g(\Delta)+m^2 h(\Delta)+j(\Delta)\right)\right)=\\
   &=\int_0^1 dx\frac{\lambda\mu^{\epsilon} }{(2\pi)^{4-\epsilon}}(\Delta^{\mu}\Delta^{\nu}-
   \Delta^2g^{\mu\nu})(1-x)x\pi^{2-\frac{\epsilon}{2}}\frac{\Gamma(\frac{\epsilon}{2})}{\overline{m}(x,t)^{\epsilon}}
   +g^{\mu \nu}\frac{\lambda\mu^{\epsilon}}{2(4\pi)^{2-\frac{\epsilon}{2}}}\frac{\Gamma(\frac{\epsilon}{2}-1)}{m^{\epsilon-2}}\\
   &\xrightarrow[]{\epsilon \to 0}\frac{\lambda}{(4\pi)^2}\int_0^1 dx(\Delta^{\mu}\Delta^{\nu}-
   \Delta^2g^{\mu\nu})(1-x)x\biggl(\frac{2}{\epsilon}+\ln{\frac{\mu^2}{\overline{m}^2(x,t)}}\biggr)+\frac{g^{\mu\nu}m^2\lambda}{2(4\pi)^2}\biggl(\frac{2}{\epsilon}-\gamma+\log{\frac{4\pi\mu^2}{m^2}}\biggr)\,.
 \end{align*}
Thus, the canonical EMT is
\begin{align}
    \bra{\Omega}T(a_{p'}T_{\text{can}}^{\mu \nu}(0)a_p^{\dagger})\ket{\Omega}
    &=\frac{P^{\mu}P^\nu}{2}-\frac{\Delta^{\mu}\Delta^\nu-g^{\mu \nu}\Delta^2}{6}
    +\frac{\lambda}{(4\pi)^2}\int_0^1 dx(\Delta^{\mu}\Delta^{\nu} -
    \Delta^2g^{\mu\nu})(1-x)x\biggl(\frac{2}{\epsilon}+\ln{\frac{\mu^2}{\overline{m}^2(x,t)}}\biggr)\nonumber\\
    &\hspace{4.3cm}
     +\frac{\lambda}{(4\pi)^2}\;\frac{g^{\mu\nu}m^2}{2}\biggl(\frac{2}{\epsilon}-\gamma+\log{\frac{4\pi\mu^2}{m^2}}\biggr)+g^{\mu\nu}\delta_m
     \nonumber\\
    &=\frac{P^{\mu}P^\nu}{2}-\frac{\Delta^{\mu}\Delta^\nu-g^{\mu \nu}\Delta^2}{2}+\frac{\lambda}{(4\pi)^2}\int_0^1 dx(\Delta^{\mu}\Delta^{\nu}-
    \Delta^2g^{\mu\nu})(1-x)x\biggl(\frac{2}{\epsilon}+\ln{\frac{\mu^2}{\overline{m}^2(x,t)}}\biggr).\label{canemt}
\end{align}
From this result we read off the contribution from the canonical EMT to the EMT form factors in dimensional regularization
quoted in Eq.~(\ref{Eq:EMT-can-dim-result}).

\subsection{Evaluation of the improvement term in dimensional regularization}
\label{App:C}

The improvement term consists of two terms which we will calculate each term individually 
\begin{gather*}
    -h\bra{\Omega}T(a_{p'}(\partial^{\mu}\partial^{\nu}-g^{\mu \nu}\Box)\Phi^2 a_p^{\dagger})\ket{\Omega}
    =-h\biggl(\bra{\Omega}T(a_{p'}\partial^{\mu}\partial^{\nu}\Phi^2 a_p^{\dagger})\ket{\Omega}-g^{\mu \nu}\bra{\Omega}T(a_{p'}\Box \Phi^2 a_p^{\dagger})\ket{\Omega}\biggr).
\end{gather*}
Note that $h=\frac{1}{4}\frac{(N-2)}{(N-1)}=\frac{1}{6}$, 
with $N=4$ being the number of space-time dimensions. It is important to iterate that $h$ is just a 
constant that we set to $N=4$ based on the number of dimensions we are working in. 
When we use dimensional regularization, we only assume the dimension $d=4-\epsilon$ of the integral
itself varies rather than the dimension $N=4$ of the manifold (physical space) we are working on,
see the discussion at the end of Sec.~\ref{Sec-3.2:EMT-dim-reg-improve}. 

For the first term we obtain
\begin{align*}
    &\hspace{-2cm}
    -h\bra{\Omega}T(a_{p'}\partial^{\mu}\partial^{\nu}\Phi(x)\Phi(x) a_p^{\dagger})\ket{\Omega}
    \\
    &=-h\partial^{\mu}\partial^{\nu}\biggl[2e^{i\Delta \cdot x}
    -i\mu^{4-d}\lambda \int d^4y e^{i\Delta\cdot y}\int\frac{d^4k}{(2\pi)^4}\frac{ie^{-ik\cdot (x-y)}}{k^2-m^2+i\epsilon}\int\frac{d^4q}{(2\pi)^4}\frac{ie^{-iq \cdot (y-x)}}{q^2-m^2+i\epsilon}\biggr] \\
    &=\biggl[2 h\Delta^{\mu}\Delta^{\nu}+L^{\mu\nu}(\Delta)\biggr]e^{i\Delta \cdot x},
\end{align*}
where
 \begin{align*}
    L^{\mu\nu}(\Delta)
    &=hi\mu^{4-d}\lambda \int\frac{d^dk}{(2\pi)^d}\frac{\Delta^{\mu}\Delta^{\nu}}{k^2-m^2+i\epsilon}
    \frac{1}{(\Delta+k)^2-m^2+i\epsilon}\\
    &=hi\mu^{4-d}\lambda \int \frac{d^dk \Delta^{\mu}\Delta^{\nu}}{(2\pi)^d}\int_0^1 dx
    \frac{1}{((k+\Delta x)^2-\overline{m}^2(x,t)+i\epsilon)^2}\\
    &=h\mu^{4-d}i\lambda \int \frac{d^d k \Delta^{\mu}\Delta^{\nu}}{(2\pi)^d}\int_0^1 dx
    \frac{1}{(k^2-\overline{m}^2(x,t)+i\epsilon)^2}\\
    &=-h\mu^{4-d}\lambda \int_0^1 dx \frac{\Delta^{\mu}\Delta^{\nu}}{(4\pi)^{\frac{d}{2}}}\Gamma\bigl(2-\frac{d}{2}\bigr)\overline{m}(x,t)^{(d-4)}.
 \end{align*}
The second term is just the trace of the first term multiplied by $g^{\mu\nu}$, i.e.\ 
 \begin{align*}
     &h g^{\mu \nu}\bra{\Omega}T(a_{p'} \Box \Phi^2 a_p^{\dagger})\ket{\Omega}
     =-g^{\mu \nu}\biggl[2h\Delta^2 + L^\rho_{\;\rho}(\Delta)\biggr]e^{i\Delta\cdot x} 
 \end{align*}

Combining terms and setting $d=4-\epsilon$ gives the result
 \begin{align*}
     &-h\bra{\Omega}T(a_{p'}(\partial^{\mu}\partial^{\nu}-g^{\mu \nu}\Box)\Phi^2 a_p^{\dagger})\ket{\Omega}
     =2h(\Delta^{\mu}\Delta^{\nu}-g^{\mu\nu}\Delta^2)\biggl[1- 
     \frac{\mu^{\epsilon}\lambda}{2(4\pi)^{\frac{\epsilon}{2}}}\,\Gamma\bigl(2-\frac{d}{2}\bigr)
     \int_0^1 dx \,\overline{m}(x,t)^{-\epsilon}\biggr]\,.
 \end{align*}
Expanding this expression in $\epsilon$ yields the result for the
improvement term contribution to the EMT quoted in Eq.~(\ref{Eq:EMT-imp-dim-result}).

\subsection{Total EMT in dimensional regularization}
\label{App:Cnew}

Combining the results from App.~\ref{App:B} and App.~\ref{App:C} and expanding in 
$\epsilon$ it follows,
 \begin{align*}
     &\hspace{-0.5cm}\bra{\Omega}T(a_{p'}(T^{\mu \nu}(0)_{\rm can}+\Theta^{\mu \nu}_{\rm imp}(0))a_p^{\dagger})\ket{\Omega}
     =\frac{P^{\mu}P^\nu }{2}-\frac{\Delta^{\mu}\Delta^\nu -g^{\mu \nu}\Delta^2}{2}
     +2h(\Delta^{\mu}\Delta^{\nu}-g^{\mu\nu}\Delta^2)+\\
    &\hspace{2cm}+\frac{\lambda}{(4\pi)^2}\int_0^1 dx(\Delta^{\mu}\Delta^{\nu}-
   \Delta^2g^{\mu\nu})\biggl[(1-x)x\biggl(\frac{2}{\epsilon}+\log{\frac{\mu^2}{\overline{m}^2(x,t)}}\biggr)-\frac{1}{6}\biggl(\frac{2}{\epsilon}+\log{\frac{\mu^2}{\overline{m}^2(x,t)}}\biggr)\biggr]\\
   &=\frac{P^{\mu}P^\nu }{2}-\frac{1}{3}\frac{\Delta^{\mu}\Delta^\nu -g^{\mu \nu}\Delta^2}{2}
    +\frac{\lambda}{(4\pi)^2}\int_0^1 dx(\Delta^{\mu}\Delta^{\nu}-
   \Delta^2g^{\mu\nu})\biggl(\frac{1}{6}-(1-x)x\biggr)\log{\frac{m^2\bigl(1-\frac{\Delta^2}{m^2}x(1-x)\bigr)}{\mu^2}}\\
   &=\frac{P^{\mu}P^\nu }{2}-\frac{1}{3}\frac{\Delta^{\mu}\Delta^\nu -g^{\mu \nu}\Delta^2}{2}
    +\frac{\lambda}{(4\pi)^2}\int_0^1 dx(\Delta^{\mu}\Delta^{\nu}-
   \Delta^2g^{\mu\nu})\biggl(\frac{1}{6}-(1-x)x\biggr)\log{\biggl(1-\frac{\Delta^2}{m^2}x(1-x)\biggr)}.
 \end{align*}

Here it should be noted that explicit dependence on $\mu$ drops 
out since $\mu$ can be separated from the original logarithm and 
$\int_0^1 dx \bigl(\frac{1}{6}-x(1-x)\bigr)\log{\frac{m^2}{\mu^2}}=0$. 
The vanishing of the integral $\int_0^1 \bigl(\frac{1}{6}-x(1-x)\bigr)$ 
ensures also the finiteness of the trace which in turn ensures the 
finiteness of the EMT \cite{Callan:1970ze}.

\section{Pauli Villars Regularization} 
\label{App:D}

In this Appendix we present details of the calculation in the Pauli-Villars regularization.
It is convenient to define 
\begin{align*}
    &M_1^2(\Delta)=\Lambda_1^2-\Delta^2x(1-x), \quad M_2^2(\Delta)=\Lambda_2^2-\Delta^2x(1-x),
\end{align*}
and for brevity below $\overline{m}=\overline{m}(x,t)$. The results for 
$F^{\mu\nu}(\Delta)$, $G^{\mu\nu}(\Delta)$, $H^{\mu\nu}(\Delta)$, $I^{\mu\nu}(\Delta)$
in Eq.~\eqref{PV} follow from 
\begin{align*}
    &
    \bra{\Omega}T(a_{p'}(\partial^{\mu}\Phi\partial^{\nu}\Phi-\partial^{\mu}\Phi_1\partial^{\nu}\Phi_1-\partial^{\mu}\Phi_2\partial^{\nu}\Phi_2) a_p^{\dagger})\ket{\Omega} \ph \\
    &=\biggl[p'^{\mu}p^{\nu}+p'^{\nu}p^{\mu}
    +i\lambda \int\frac{d^4k}{(2\pi)^4}\int_0^1 dx \biggl(\frac{g^{\mu\nu}}{4}k^{2}-\Delta^{\mu}\Delta^{\nu}x+\Delta^{\mu}\Delta^{\nu}x^2\biggr)\times\\
    &\hspace{1cm}\biggl(
    \frac{1}{(k^2-\overline{m}^2(x,t)+i\epsilon)^2}- \frac{a_1^2}{(k^2-M_1^2(\Delta)+i\epsilon)^2}- \frac{a_2^2}{(k^2-M_2^2(\Delta)+i\epsilon)^2}\biggr)\biggr]\\
    & \hspace{2cm}\xrightarrow[]{\Lambda_1^2,\Lambda_2^2\to \Lambda^2 \gg \overline{m}^2}\biggl[p'^{\mu}p^{\nu}+p'^{\nu}p^{\mu}+\int_0^1 dx \frac{\lambda}{(4\pi)^2}\biggl(\frac{ g^{\mu\nu}}{2}\overline{m}^2\log{\frac{\overline{m}^2}{\Lambda^2}}+\frac{\Delta^{\mu}\Delta^{\nu}x(1-x)}{2}\log{\frac{\Lambda^2}{\overline{m}^2}}\biggr)\biggr]=F^{\mu\nu}(\Delta), \\
    &
    -\frac{g^{\mu\nu}}{2}\bra{\Omega}T(a_{p'}(\partial^{\mu}\Phi\partial_{\mu}\Phi-\partial^{\mu}\Phi_1\partial_{\mu}\Phi_1-\partial^{\mu}\Phi_2\partial_{\mu}\Phi_2) a_p^{\dagger})\ket{\Omega}=\\
    &=\biggl[g^{\mu\nu} p'\cdot p
    -i\lambda g^{\mu\nu} \int\frac{d^4k}{2(2\pi)^4}\int_0^1 dx (k^{2}-\Delta^{2}x(1-x))\times\\
    &\hspace{1cm}\biggl(
    \frac{1}{(k^2-\overline{m}^2(x,t)+i\epsilon)^2}- \frac{a_1^2}{(k^2-M_1^2(\Delta)+i\epsilon)^2}- \frac{a_2^2}{(k^2-M_2^2(\Delta)+i\epsilon)^2}\biggr)\biggr]\\
    &\hspace{2cm}\xrightarrow[]{\Lambda_1^2,\Lambda_2^2\to \Lambda^2 \gg \overline{m}^2}-\biggl[ g^{\mu\nu} p'\cdot p+\int_0^1 dx\frac{\lambda}{(4\pi)^2}\biggl( g^{\mu\nu}\overline{m}^2\log{\frac{\overline{m}^2}{\Lambda^2}}+\frac{g^{\mu\nu}\Delta^{2}x(1-x)}{2}\log{\frac{\Lambda^2}{\overline{m}^2}}\biggr)\biggr]=G^{\mu\nu}(\Delta) \, , \\ 
   &
   \frac{g^{\mu\nu}}{2}\bra{\Omega}T(a_{p'}((m^2+\delta_m) \Phi\Phi-\Lambda_1^2\Phi_1\Phi_1-\Lambda_2^2\Phi_2\Phi_2)a_{p}^{\dagger})\ket{\Omega}]\\
    &=g^{\mu\nu}\biggl[m^2+\delta_m-\frac{m^2}{2}i\lambda\int_0^1 \frac{d^4k}{(4\pi)^2}\int dx\biggl(
    \frac{1}{(k^2-\overline{m}^2(x,t)+i\epsilon)^2}- \frac{a_1^2}{(k^2-M_1^2(\Delta)+i\epsilon)^2}- \frac{a_2^2}{(k^2-M_2^2(\Delta)+i\epsilon)^2}\biggr)\biggr]\\
    &\hspace{2cm}
   \xrightarrow[]{\Lambda_1^2,\Lambda_2^2\to \Lambda^2 \gg \overline{m}^2}
    g^{\mu\nu}\biggl(m^2+\delta_m-\int_0^1 dx\frac{\lambda}{2(4 \pi)^2}m^2\log{\frac{\Lambda^2}{\overline{m}^2}}\biggr)=H^{\mu\nu}(\Delta)+g^{\mu\nu}\delta_m,\\ 
  &\frac{\lambda}{4!}\bra{\Omega}T(a_{p'}(\Phi+a_1\Phi_1+a_2\Phi_2)^4 a_{p}^{\dagger}e^{-i\int \frac{\lambda}{4!}\Phi^4d^4x})\ket{\Omega}
   =\frac{\lambda}{2(4\pi)^2}\biggl(\Lambda^2+m^2\log{\frac{\Lambda^2}{m^2}}\biggr).
\end{align*}   
Similarly, for the improvement term it follows
\begin{align*}
    &-h\bra{\Omega}T(a_{p'}(\partial^{\mu}\partial^{\nu}\Phi\Phi -\partial^{\mu}\partial^{\nu}\Phi_1\Phi_1-\partial^{\mu}\partial^{\nu}\Phi_2\Phi_2)a_p^{\dagger})\ket{\Omega}
    +h g^{\mu \nu}\bra{\Omega}T(a_{p'}(\Box \Phi^2 -\Box \Phi_1^2-\Box \Phi_2^2)a_p^{\dagger})\ket{\Omega}\\
    &=(\Delta^{\mu}\Delta^{\nu}-g^{\mu\nu}\Delta^2)\biggl(2h-h\frac{\lambda}{(4\pi)^2}\log{\frac{\Lambda^2}{\overline{m}^2}}\biggr)=I^{\mu\nu}(\Delta).
\end{align*}
Combining the results yields for the total EMT
\begin{align*}
     &\hspace{-0.5cm}
     \bra{ \vec{ p'}} (T^{\mu\nu}(0)_{\rm can}+\Theta^{\mu\nu}(0)_{\rm imp}) \ket{ \vec {p}}
     =\frac{P^{\mu}P^\nu }{2}-\frac{1}{3}\frac{\Delta^{\mu}\Delta^\nu -g^{\mu \nu}\Delta^2}{2}\\
     &\hspace{1cm}+\frac{\lambda}{(4 \pi)^2}\int_0^1 dx\biggl(\frac{g^{\mu\nu}}{2}\bar{m}^2\log{\frac{\bar{m}^2}{\Lambda^2}}+\Delta^{\mu}\Delta^{\nu}x(1-x)\log{\frac{\Lambda^2}{\bar{m}^2}}- g^{\mu\nu}\bar{m}^2\log{\frac{\bar{m}^2}{\Lambda^2}}
    -\frac{g^{\mu\nu}\Delta^{2}x(1-x)}{2}\log{\frac{\Lambda^2}{\bar{m}^2}}\\
    &\hspace{1cm}+g^{\mu\nu}\biggl(\delta_m-\frac{1}{2}m^2\log{\frac{\Lambda^2}{\bar{m}^2}}\biggr)+\frac{1}{2}\biggl(\Lambda^2+m^2\log{\frac{\Lambda^2}{m^2}}\biggr)
    -h(\Delta^{\mu}\Delta^{\nu}-g^{\mu\nu}\Delta^2)\log{\frac{\Lambda^2}{\bar{m}^2}}\biggr)\\
    &=\frac{P^{\mu}P^\nu }{2}-\frac{1}{3}\frac{\Delta^{\mu}\Delta^\nu -g^{\mu \nu}\Delta^2}{2}+(\Delta^{\mu}\Delta^\nu -g^{\mu \nu}\Delta^2)\frac{\lambda}{(4\pi)^2}\int_0^1 dx\biggl((1-x)x-\frac{1}{6}\biggr)\log{\frac{\Lambda^2(1-\frac{\Delta^2x(1-x)}{\Lambda^2})}{m^2(1-\frac{\Delta^2}{m^2}x(1-x))}}\\
    & \xrightarrow[]{\Lambda\to \infty} 
    \frac{P^{\mu}P^\nu }{2}-\frac{1}{3}\frac{\Delta^{\mu}\Delta^\nu -g^{\mu \nu}\Delta^2}{2}
    +(\Delta^{\mu}\Delta^\nu -g^{\mu \nu}\Delta^2)\frac{\lambda}{(4\pi)^2}\int_0^1 dx\biggl(-(1-x)x+\frac{1}{6}\biggr)\log{\biggl(1-\frac{\Delta^2}{m^2}x(1-x)\biggr)},
\end{align*}
where in the last term, before taking $\Lambda\to\infty$,
we made use of the fact that the term proportional to 
$\log[\frac{\Lambda^2}{m^2}]$ can be separated from 
the main logarithm and drops out due to 
$\int_0^1 dx \bigl(\frac{1}{6}-x(1-x)\bigr)\log[\frac{\Lambda^2}{m^2}]=0$.
We recall that the vanishing of $\int_0^1 dx \bigl(\frac{1}{6}-x(1-x)\bigr)$
ensures the finiteness of the EMT \cite{Callan:1970ze}, see the discussion at the end
of App.~\ref{App:Cnew}.


\end{document}